\documentclass[paper]{pnastwo}

\usepackage{graphicx}
\usepackage{color}
\usepackage{epsfig}
\usepackage{epstopdf}
\usepackage{subfigure}
\contributor{Submitted to Proceedings of the National Academy of Sciences of the United States of America}
\url{www.pnas.org/cgi/doi/10.1073/pnas.0709640104}
\copyrightyear{2008}
\issuedate{Issue Date}
\volume{Volume}
\issuenumber{Issue Number}

\begin{document}

\title{Heat transport in boiling turbulent Rayleigh-B\'{e}nard convection}

\author{Rajaram Lakkaraju\affil{1}{Physics of Fluids, Faculty of Science and Technology, Mesa+ Institute and J. M. Burgers Center for Fluid Dynamics, University of Twente, 7500AE Enschede, The Netherlands},
Richard J. A. M. Stevens\affil{1}{Physics of Fluids, Faculty of Science and Technology, Mesa+ Institute, University of Twente, 7522 NB, Enschede, The Netherlands.}\affil{2}{Department of Mechanical Engineering, Johns Hopkins University, Baltimore, MD 21218, USA}, 
Paolo Oresta\affil{3}{Department of Mathematics, Mechanics and Management, Polytechnic of Bari, Via Japigia 182, 70126, Bari, Italy.}\affil{4}{Department of Engineering for Innovation, University of Salento, Via per Arnesano, 73100 Lecce, Italy.}, 
Roberto Verzicco\affil{1}{Physics of Fluids, Faculty of Science and Technology, Mesa+ Institute and J. M. Burgers Center for Fluid Dynamics, University of Twente, 7500AE Enschede, The Netherlands.}\affil{5}{Department of Mechanical Engineering, University of Rome `Tor Vergata', Via del Politecnico 1, 00133 Rome, Italy}, 
Detlef Lohse\affil{1}{Physics of Fluids, Faculty of Science and Technology, Mesa+ Institute and J. M. Burgers Center for Fluid Dynamics, University of Twente, 7500AE Enschede, The Netherlands}, 
\and
Andrea Prosperetti\affil{1}{Physics of Fluids, Faculty of Science and Technology, Mesa+ Institute and J. M. Burgers Center for Fluid Dynamics, University of Twente, 7500AE Enschede, The Netherlands.}\affil{2}{Department of Mechanical Engineering, Johns Hopkins University, Baltimore, MD 21218, USA}
}

\contributor{Submitted to Proceedings of the National Academy of Sciences of the United States of America}

\maketitle

\begin{article}
\begin{abstract} 

Boiling is an extremely effective way to promote heat transfer from a hot 
surface to a liquid due to several mechanisms many of which are not 
understood in quantitative detail. An important component of the overall 
process is that the buoyancy of the bubbles compounds with that of the 
liquid to give rise to a much enhanced natural convection. In this paper 
\textcolor{black}{we focus specifically on this enhancement and} present a numerical 
study of \textcolor{black}{the resulting} two-phase Rayleigh-B\'{e}nard convection 
process. \textcolor{black}{We make no attempt to model other aspects of the boiling 
process such as bubble nucleation and detachment.} 
We consider a cylindrical cell with a diameter equal to its height. 
The cell base and top are held at temperatures above and below the boiling 
point of the liquid, respectively. By keeping the temperature difference 
constant and changing the liquid pressure we study the effect of the 
liquid superheat in a Rayleigh number range that, in the absence of boiling, 
would be between $2\times10^6$ and $5\times10^9$. We find a considerable 
enhancement of the heat transfer and study its dependence on the bubble 
number, the degree of superheat of the hot cell bottom and the Rayleigh 
number. The increased buoyancy provided by the bubbles leads to more energetic 
hot plumes detaching from the hot cell bottom and, as a consequence, the 
strength of the circulation in the cell is significantly increased. Our 
results are in general agreement with recent experimental results of Zhong 
et al., Phys. Rev. Lett. {\bf 102}, 124501 (2009) for boiling 
Rayleigh-B\'{e}nard convection.

\end{abstract}


The greatly enhanced heat transfer brought about by the boiling process
is believed to be due to several interacting components, see, e.g., \cite{dhir98,carey1992,lienhard2011}. 
With their growth the bubbles cause 
a micro-convective motion on the heating surface and, as they detach by 
buoyancy, the volume they vacate tends to be replaced by cooler liquid. 
Especially in subcooled conditions, the liquid in the relatively stagnant 
microlayer under the bubbles can evaporate and condense on the cooler bubble 
top. This process provides for the direct transport of latent heat, which is 
thus able to bypass the low-velocity liquid region adjacent to  the heated 
surface due by the no-slip condition. The bubble growth process itself requires 
latent heat and, therefore, also removes heat from the heated surface and the 
neighboring hot liquid. Finally, with their buoyancy, the bubbles enhance the 
convective motion in the liquid beyond the level caused by the well-known 
single-phase Rayleigh-B\'{e}nard convection 
mechanisms, see e.g.,~\cite{dhir98,carey1992,lienhard2011}. This \textcolor{black}{last 
process} is the aspect on which we focus in the present paper. 

\begin{figure}[h!]
 \label{fig:Nuphaseplot}
\begin{center}
 \includegraphics[width=0.5\textwidth]{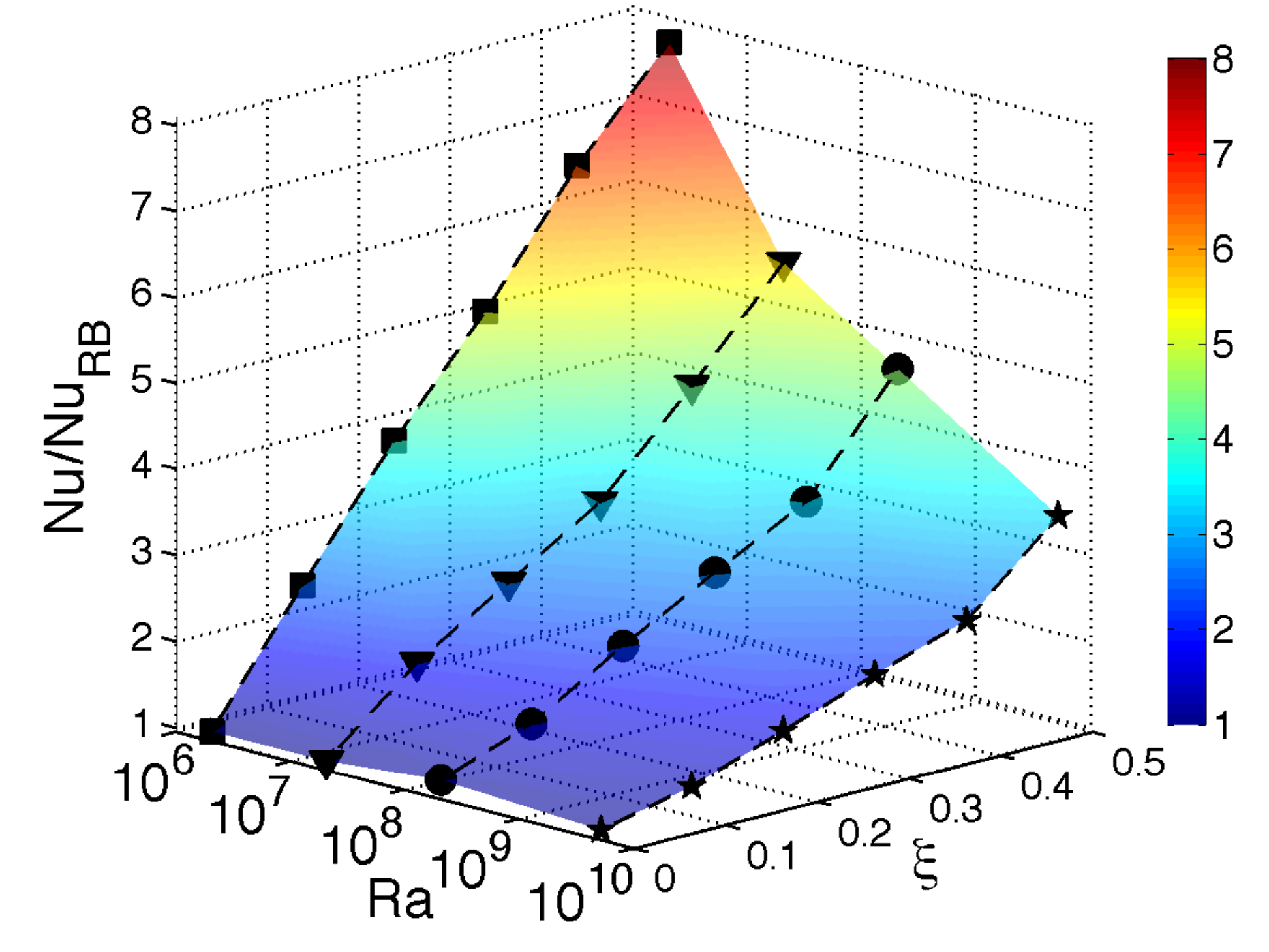} 
  \caption{$Nu(Ra, \xi)$ for boiling convection normalized by the corresponding single-phase value $Nu_{RB}$ for $N_b=50000$ bubbles. Here $\xi$ is the normalized superheat, $\xi \equiv {(T_h-T_{sat})/ \Delta}$. The symbols correspond to $Ra=2\times 10^6$ (square), $2\times 10^7$ (triangles), $Ra=2\times 10^8$ (circles) and $5\times 10^9$ (stars).}
\end{center}
\end{figure}

In classical single-phase Rayleigh-B\'{e}nard (RB) convection, the dimensionless heat transport, $Nu$, the Nusselt number, is defined as the ratio of the total heat transported through the cell to the heat that would be transported by pure conduction with a quiescent fluid. This ratio increases well above 1 as the Rayleigh number $Ra={{g\beta\Delta L^3}\over{\nu\kappa}}$ is increased due to the onset of circulatory motion in the cell. Here $g$ is the acceleration of gravity, $\beta$ the isobaric thermal expansion coefficient, $\Delta=T_h-T_c$ the difference between the temperature $T_h$ of the hot bottom plate and the temperature $T_c$ of the cold top plate, $L$ the height of the cell, $\nu$ the kinematic viscosity and $\kappa$ the thermal diffusivity. 
Further, $Nu$ depends on the shape of the cell, its aspect ratio (defined for a cylindrical cell of diameter $D$ as $\Gamma=D/L$) and the Prandtl number $Pr=\nu/\kappa$ of the liquid. 
For $Ra$ in the range $10^7-10^{10}$ and $Pr$ in the range $0.7-7$, the heat transport satisfies an approximate scaling relation $Nu\propto Ra^{0.29-0.32}$~\cite{lohsermp,lohsearfm}. 

How is this scaling modified if the hot plate temperature $T_h$ is above the fluid saturation temperature $T_{sat}$, so that \textcolor{black}{phase change can occur}? 
\textcolor{black}{The present paper addresses this question focusing on the enhanced 
convection caused by the bubble buoyancy, rather than attempting a 
comprehensive modeling of the actual boiling process in all its complexity.}
We carry out numerical simulations in the range $2\times 10^6\le Ra\le 5\times 10^9$ for a cylindrical cell with aspect ratio $\Gamma=1$ for $Pr=1.75$, which is appropriate for water at 100 $^o$C under normal conditions.

\textcolor{black}{This work differs in two major respects from our earlier studies of 
the problem. In the first place, we are now able to reach a much higher 
Rayleigh number, 5$\times 10^9$ as opposed to 2$\times10^5$ as 
in\cite{oresta09}, and to include three times as many bubbles. 
Secondly, we now study the effect of the liquid superheat which was held
fixed before.}

The extensive literature on boiling leads to the expectation that the appearance of bubbles would cause a substantial increase in $Nu$ with respect to single-phase convection, see e.g., Ref.~\cite{dhir98}. For RB convection, the effect of phase change has recently been studied in Ref.~\cite{zhong09} for the case of ethane near the critical point, and indeed \textcolor{black}{a major increase of the}
 heat transport has been found.

\section{Model}
The present paper \textcolor{black}{is based on the same mathematical model and numerical
method that we have used in  Ref.~\cite{oresta09} and several other recent 
papers~\cite{Laura11,Lakkaraju11,Lakkaraju12} to which the reader is referred for details.} 
\textcolor{black}{Briefly, for the purposes of their interaction with the liquid, the 
bubbles are modelled as point sources of momentum and heat for the liquid 
treated in the Boussinesq approximation. The motion of each bubble, envisaged 
as a sphere, is followed 
in a Lagrangian way by means of an equation which, in addition to 
buoyancy, includes drag, added mass and lift. In its mechanical aspects, 
therefore, the model is similar to existing ones which have been extensively 
used in the literature to simulate dilute disperse flows with bubbles and 
particles (see e.g. Refs.~\cite{prosperetti2004,prosperetti2007}). The novelty of our model lies in the 
addition of the thermal component. The heat exchange between the bubble and 
the liquid in its vicinity is modelled by means of a heat transfer coefficient 
dependent on the Peclet number of the bubble-liquid relative motion and 
on the Prandtl number of the liquid. The radial motion of the bubbles is slow 
enough that the vapor pressure remains essentially equal to the ambient 
pressure, which implies that the bubble surface temperature can be assumed to 
remain at the saturation value. The bubble volume, on which the enhanced
buoyancy effect depends, is calculated by assuming that the entire heat 
absorbed by a bubble is used to generate vapor at the saturation density 
and pressure.} 

\textcolor{black}{The calculation is carried out on a finite-difference grid based 
on cylindrical coordinates. The standard staggered-grid arrangement is 
used for the flow variables and the projection method for the calculation 
of the pressure and time stepping. No-slip conditions are applied on the 
bottom and top of the cell and free-slip is allowed on the lateral boundary. 
The Lagrangian treatment of the bubbles proceeds by means of a third-order 
Runge-Kutta method. The energy and force imparted by each 
bubble to the liquid are interpolated to the grid points of the cell 
containing the bubble in such a way as to preserve the total energy and 
the resultant and moment of the force.}

\begin{figure}[h!]
\begin{center}
\subfigure{\includegraphics[width=0.23\textwidth]{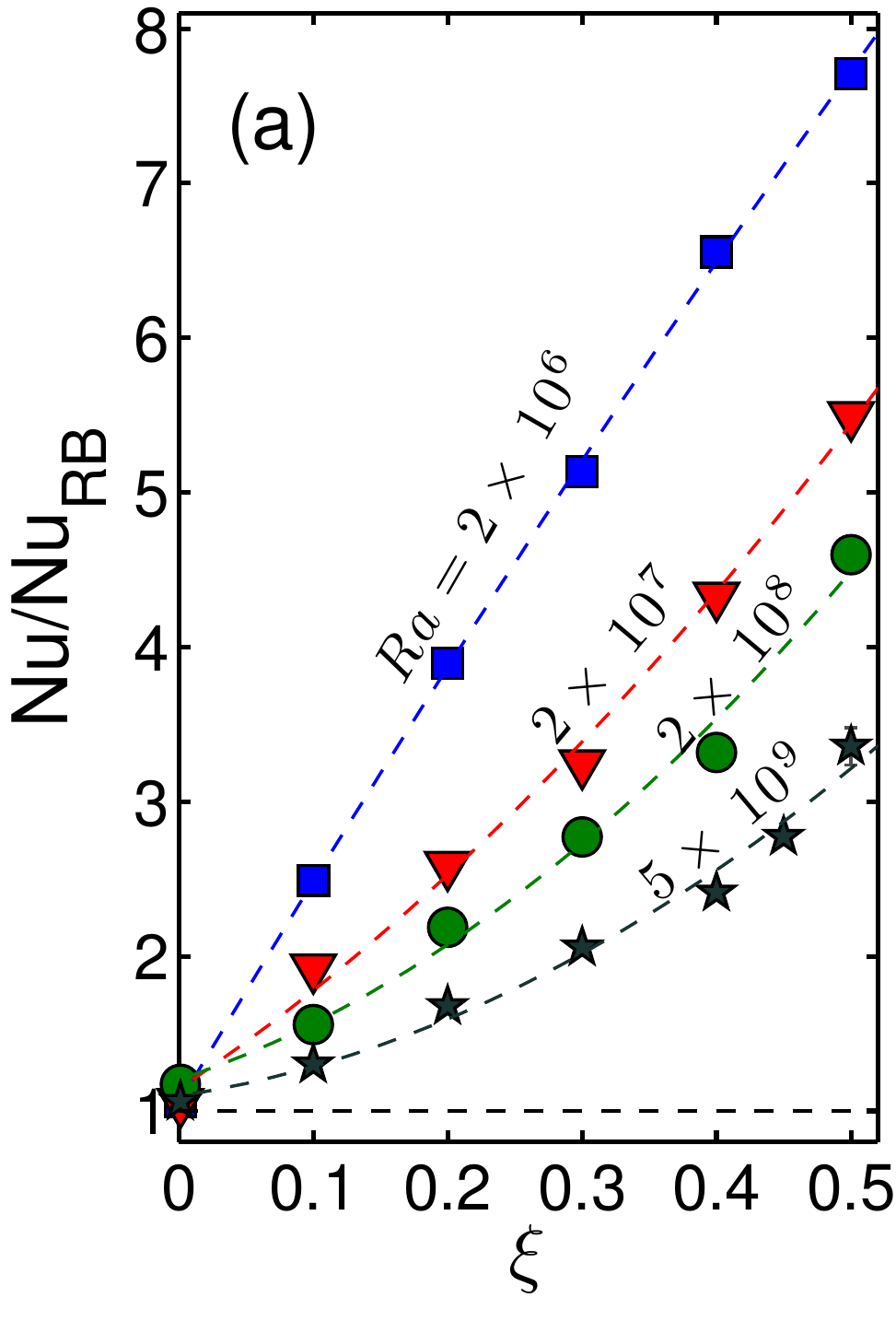}}
\subfigure{\includegraphics[width=0.23\textwidth]{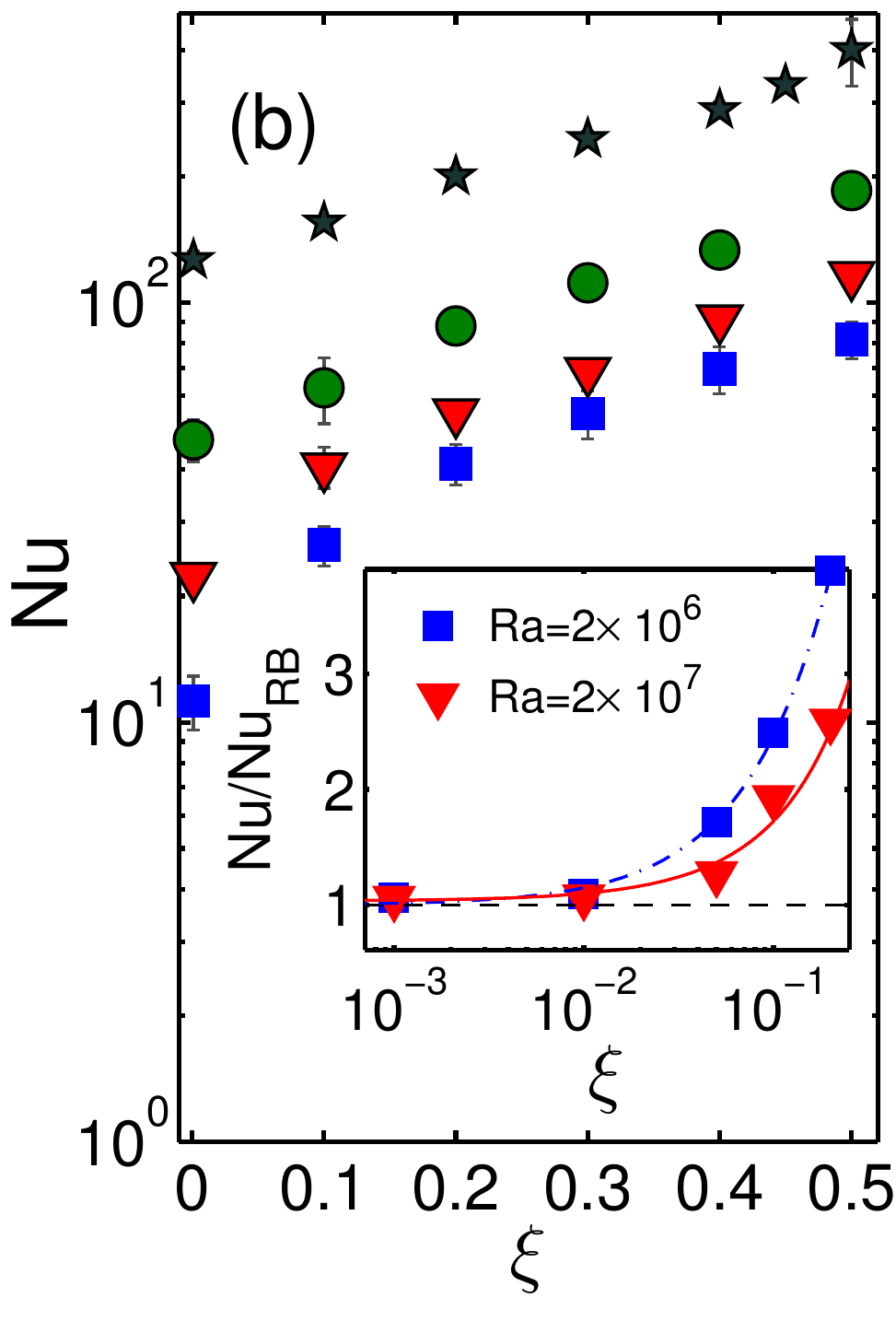}} 
\caption{$Nu/Nu_{RB}$ (left) and $Nu$ (right) as functions of the normalized superheat $\xi$ for 50000 bubbles.  The symbols correspond to $Ra=2\times10^6$ (squares), $Ra=2\times 10^7$ (triangles), $Ra=2\times10^8$ (circles) and $Ra=5\times 10^9$ (stars). The inset shows a detail for small superheat $\xi$ for $Ra=2\times 10^6$ (blue squares) and $Ra=2\times10^7$ (red triangles) with quadratic fits to the data.}
\label{fig:NuvsXi_fixedNb50}
\end{center}
\end{figure}

Simulations are carried out on computational grids with the angular, radial, and axial directions discretized by means of $193\times49\times129$, $385\times129\times257$, $385\times129\times257$, and $769\times193\times385$ nodes for $Ra=2\times10^6$, $2\times10^7$, $2\times10^8$, and $5\times10^9$. The simulations are therefore well resolved according to the requirements specified in Refs.~\cite{richard10,Lakkaraju12}. We have also checked the global balances of appendix B in Ref.~\cite{oresta09} finding that they were satisfied to within $0.1\%$.

\textcolor{black}{When a bubble reaches the top cold plate it is removed from the 
calculation to model condensation and a new bubble is introduced at a random 
positions on the bottom hot plate so that the total number of bubbles in the 
calculation remains constant. We do not attempt to model the nucleation 
process which, with the present state of knowledge, cannot be done on the 
basis of first principles and which would require addressing extremely 
complex multi-scale issues. For our limited purpose of studying the 
bubble-induced increased buoyancy it is sufficient to simply generate a 
new bubble at the hot plate. We do not model the process by which the bubble 
detaches from the plate but assume that it is free to rise immediately as it 
is introduced. The initial bubble radius is arbitrarily 
set at 38 $\mu$m. As shown in Ref.~\cite{Lakkaraju11}, the initial bubble size is immaterial provided it is in the range of a few tens of microns. In view of their smallness, the latent heat necessary for their generation is very small and is neglected. We show results for three values of the total number of 
bubbles $N_b$, namely $N_b=10000$, 50000, and 150000.} Another parameter we vary is the degree of superheat, $T_h-T_{sat}$, which we express in the dimensionless form $\xi=(T_h-T_{sat})/\Delta$. 

The Nusselt number shown in the following is defined as a $Nu={q^{''}_hL/(k\Delta)}$, where $k$ is the liquid thermal conductivity and $q^{''}_h$ is the heat flux into the bottom plate. This quantity differs from $q^{''}_c$, the heat flux at the upper plate, due to the removal of the bubbles that reach the top boundary\footnote{The Nusselt number shown in our previous papers~\cite{oresta09,Laura11,Lakkaraju11} are based on the average between $q^{''}_h$ and $q^{''}_c$.}.

An important new parameter introduced by the bubbles is the Jakob number 
$Ja={\rho c_p(T_h-T_{sat})\over{\rho_v h_{fg}}}=\xi \,{\rho c_p\Delta \over{\rho_v h_{fg}}}$, where $\rho$ and $\rho_v$ are the densities of liquid and vapor, $c_p$ is the liquid specific heat, and $h_{fg}$ is the latent heat for vaporization. Physically, $Ja$ expresses the balance between the available thermal energy and the energy required for vaporization. With $\Delta=1$ $^o$C, $Ja$ varies  between 0 and 1.68 as $\xi$ varies between 0 and 1/2.
For $\xi=0$, the bubbles introduced at the hot plate can only encounter liquid at saturation temperature or colder, and therefore they cannot grow but will mostly collapse. On the other hand, for $\xi=1/2$, they have significant potential for growth.

To give an impression of the physical situation corresponding to our parameter choices, we may mention that 100 $^o$C water in a 15 cm-high cylinder with an imposed temperature difference $\Delta=1$ $^o$C would correspond to $Ra\simeq 5\times10^8$. 
The Kolmogorov length scale based on the volume- and time-averaged kinetic energy dissipation 
in single-phase RB convection is 3 mm for $Ra\sim 10^6$ and 0.5 mm for $Ra\sim 10^{10}$, see e.g., Ref.~\cite{lohsearfm}, and is therefore always much larger than the initial size of the injected bubbles (i.e., 
$0.5\times10^{-3}/(38\times10^{-6})\approx 13$ times larger for the highest Rayleigh number).

\section{Observations on heat transport and flow organization}
In Figure~\ref{fig:Nuphaseplot}, the dependence of $Nu$ on the Rayleigh number $Ra$ and the dimensionless superheat $\xi$ is shown for $N_b=50000$ bubbles. Here $Nu$ is normalized by $Nu_{RB}$, the single-phase Nusselt number corresponding to the same value of $Ra$. Each symbol shows the result of a separate simulation carried out for the corresponding values of $Ra$ and $\xi$.  A  colored surface is interpolated through the computed results with the color red corresponding to $Nu/Nu_{RB}=8$ and the color blue $Nu/Nu_{RB}=1$.

The same data are shown on a two-dimensional plot of $Nu/Nu_{RB}$ vs. $\xi$ in Figure~\ref{fig:NuvsXi_fixedNb50}a for four different Rayleigh numbers in descending order; here the dashed lines are drawn as guides to the eye. It is evident that the {\it relative} enhancement of the heat transport is a decreasing function $Ra$. This statement, however,  does not apply to the {\it absolute} heat transport shown in Figure~\ref{fig:NuvsXi_fixedNb50}b, where $Nu$ is not normalized by the single-phase value. Here $Ra$ increases in ascending order, which shows that the bubbles always have a beneficial effect on the heat transport. For very small superheat the heat transport approaches the single-phase value as shown in the inset of Figure~\ref{fig:NuvsXi_fixedNb50}b.

Figures~\ref{fig:Nuphaseplot} and ~\ref{fig:NuvsXi_fixedNb50} show results calculated keeping the bubble number fixed. This procedure, therefore, does not faithfully reflect physical reality as it is well known that the number of bubbles is an increasing function of superheat. The dependence is actually quite strong, with the number of bubbles proportional to $T_h-T_{sat}$ raised to a power between 3 and 4~\cite{dhir98}. However, varying independently $N_b$ and $\xi$ permits us to investigate separately the effect of these quantities.

\begin{figure}[h!]
\begin{center}
\subfigure{\includegraphics[width=0.23\textwidth]{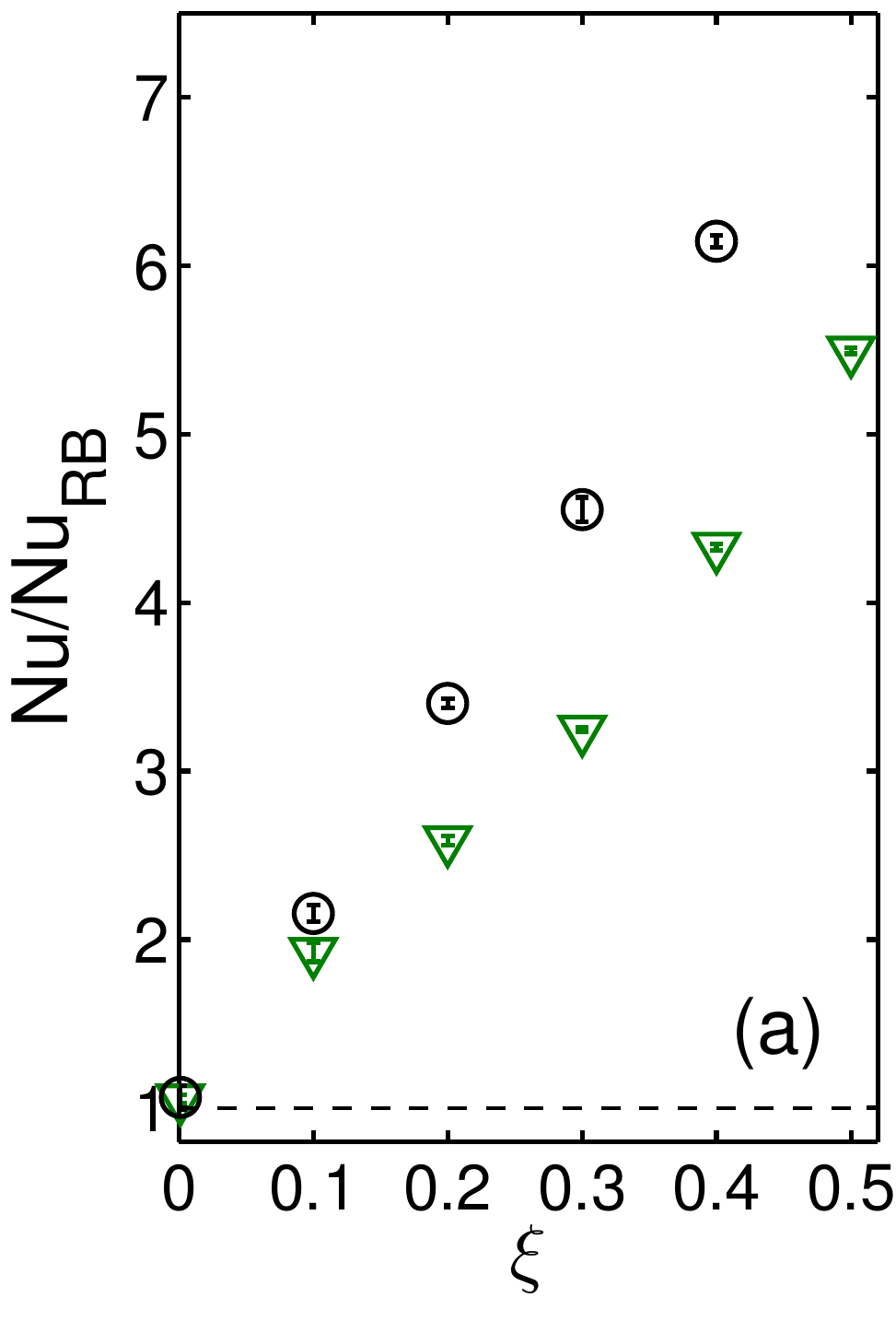}} 
\subfigure{\includegraphics[width=0.23\textwidth]{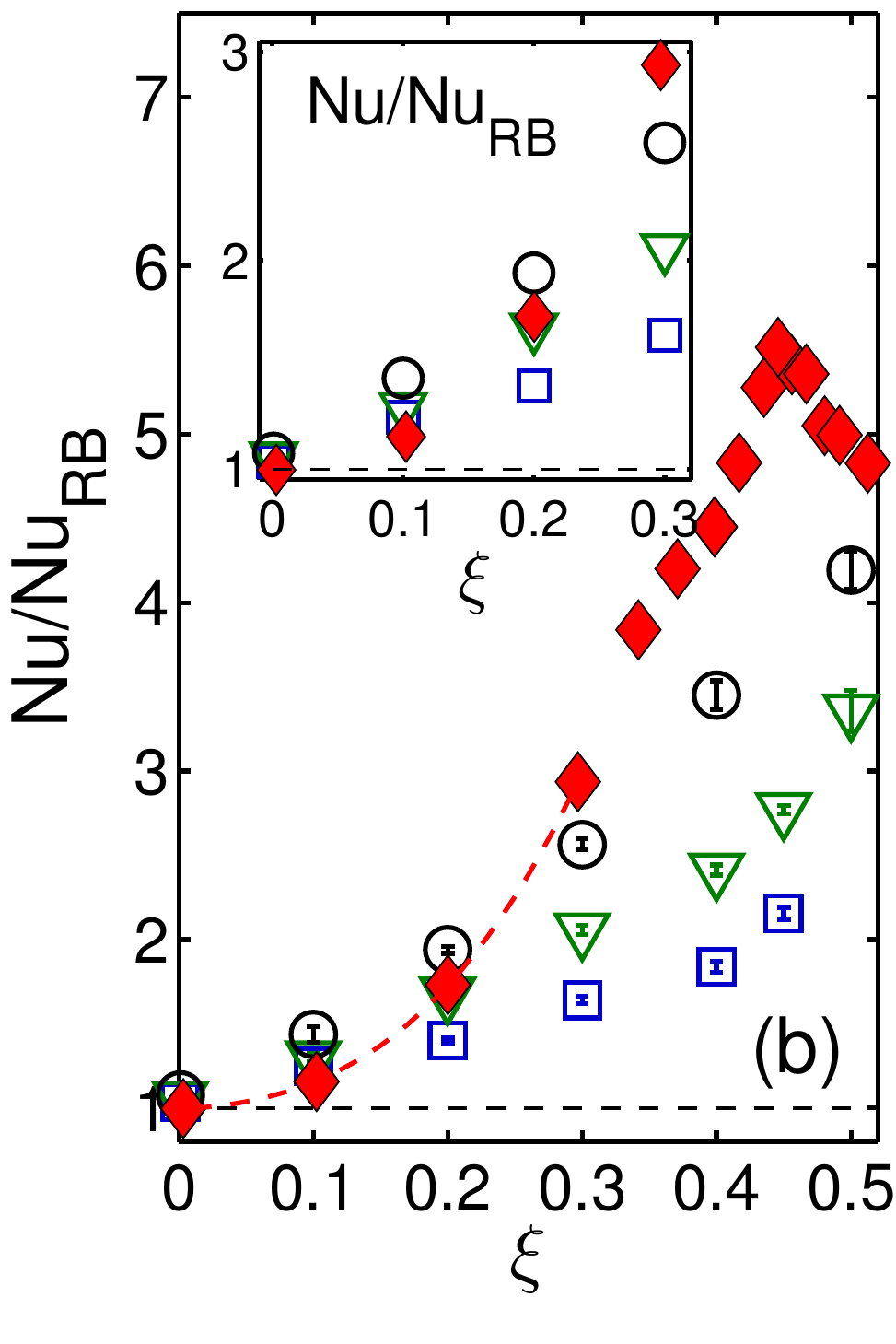}} 
\caption{$Nu/Nu_{RB}$ vs. $\xi$ for three different bubble numbers,
$N_b=10000$ (squares, right figure only), 
50000 (triangles) and 150000 (circles); the left panel is for
$Ra=2\times10^7$ and the right panel for $Ra=5\times10^9$.
The red-dashed line is a fit to the experimental data of Zhong et al.~\cite{zhong09} shown by the filled symbols.
The inset is a blow-up for the range $0\le\xi\le0.30$.}
\end{center}
\label{fig:NuvsXi_varyNb}
\end{figure}

The effect of changing the bubble number from 50000 to 150000 at the same $\xi$ is shown in Figures~\ref{fig:NuvsXi_varyNb}a and~\ref{fig:NuvsXi_varyNb}b for $Ra=2\times 10^7$ and $5\times 10^9$, respectively. In the latter case we also include results for $N_b=10000$. 
For small $\xi$ the heat transfer enhancement is small as the bubbles will mostly encounter colder liquid, condense and add very little to the system buoyancy. As the superheat $\xi$ increases, however, the effect of the bubbles become stronger and stronger, and larger the larger their number.

In Figure~\ref{fig:NuvsXi_varyNb}b, the solid symbols are the data of Ref.~\cite{zhong09} taken at a 
higher Rayleigh number, $Ra\approx 3\times 10^{10}$
\footnote{This paper reports data for both increasing and decreasing superheat. We show here only the latter data because, for increasing superheat, there is a threshold for fully developed boiling conditions which pushes the onset of bubble appearance beyond $\xi=0.35$. 
For decreasing $\xi$ on the other hand fully developed boiling conditions prevail all the way to small values of $\xi$}. 
The inset in the figure shows our computed results and the experimental data for $\xi\le0.3$. \textcolor{black}{A major difference between our simulations and the 
experiment is that, in the latter, the number of bubbles increases with 
the superheat, while it remains constant with $\xi$ in the simulations. We
can nevertheless attempt a comparison as follows.}
Quadratic interpolation using our results for the three values of $N_b$ suggests that, in order to match the experimental values, we would need $N_b\simeq$ 63000 for $\xi=$ 0.2 and 
$N_b\simeq$ 250000 for $\xi=$ 0.3. 
If, as suggested by experiment, the actual physical process results in a relation of the form $N_b\propto \xi^m$, we find $m \simeq$ 3.4 which falls in the experimental range $3<m<4$ mentioned before. With this value of $m$, we can estimate the number of bubbles necessary to account for the measured $Nu$ at $\xi=0.1$. 
Using $N_b(\xi=0.1)=(0.1/\xi)^m \, N_b(\xi)$ we find $N_b(0.1)\approx$ 5968 for $\xi=0.2$ and $N_b(0.1)\approx$6000 for $\xi=0.3$. 
These values are in agreement and consistent with the fact that our computed result at $N_b=$10000 is somewhat higher than the measured value for $\xi=0.1$. The picture that emerges from these considerations is therefore in reasonable agreement with experiment. 
A similar exercise cannot be carried out for larger values of $\xi$ as in the experiment bubbles then become so large that they coalesce and form slugs with non-negligible dimensions. Our model, \textcolor{black}{in which the vapor volume fraction is assumed to be so small as to be negligible, clearly cannot be applied to this situation.}

\begin{figure}[h!]
\begin{center}
\subfigure{\includegraphics[width=0.23\textwidth]{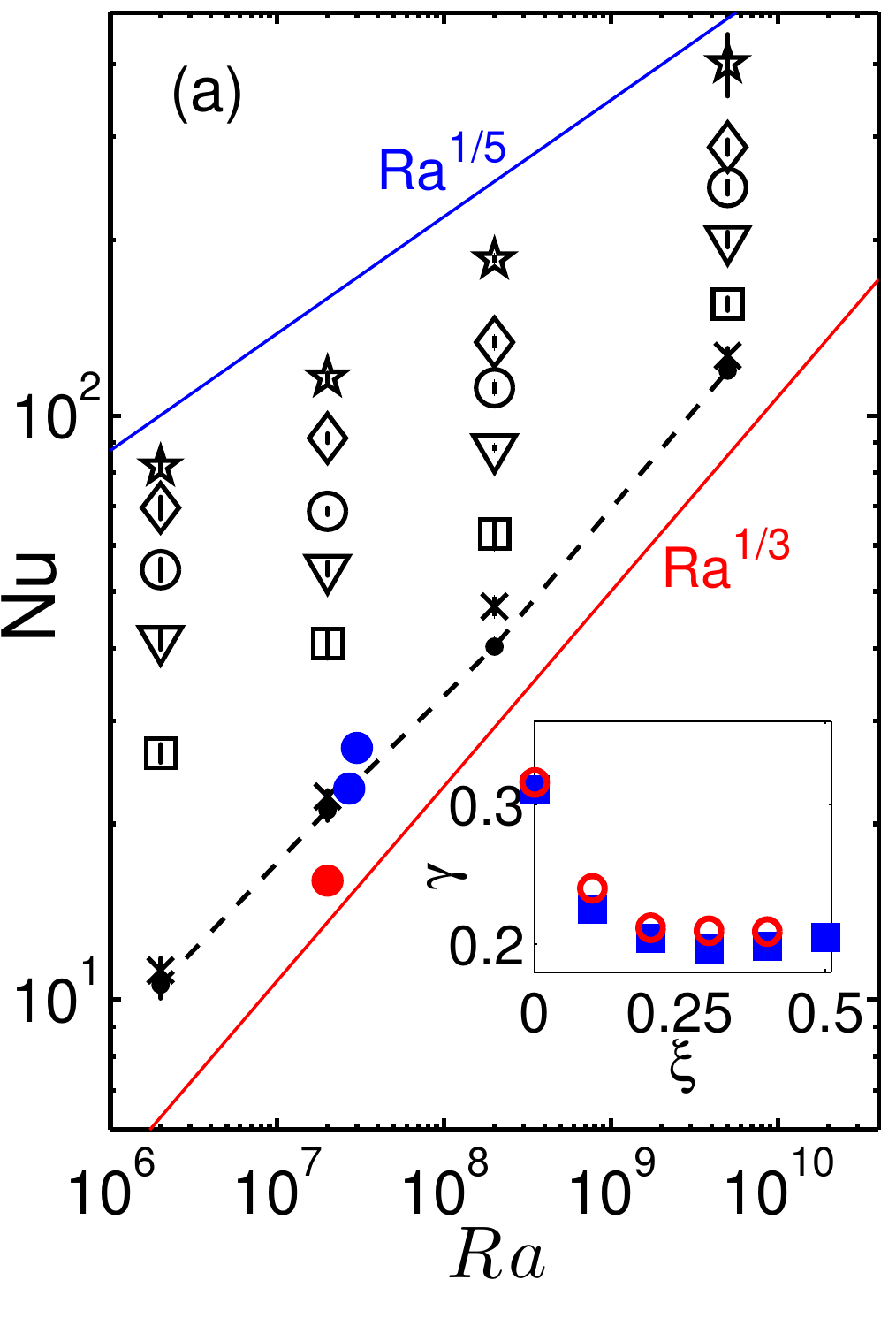}}
\subfigure{\includegraphics[width=0.23\textwidth]{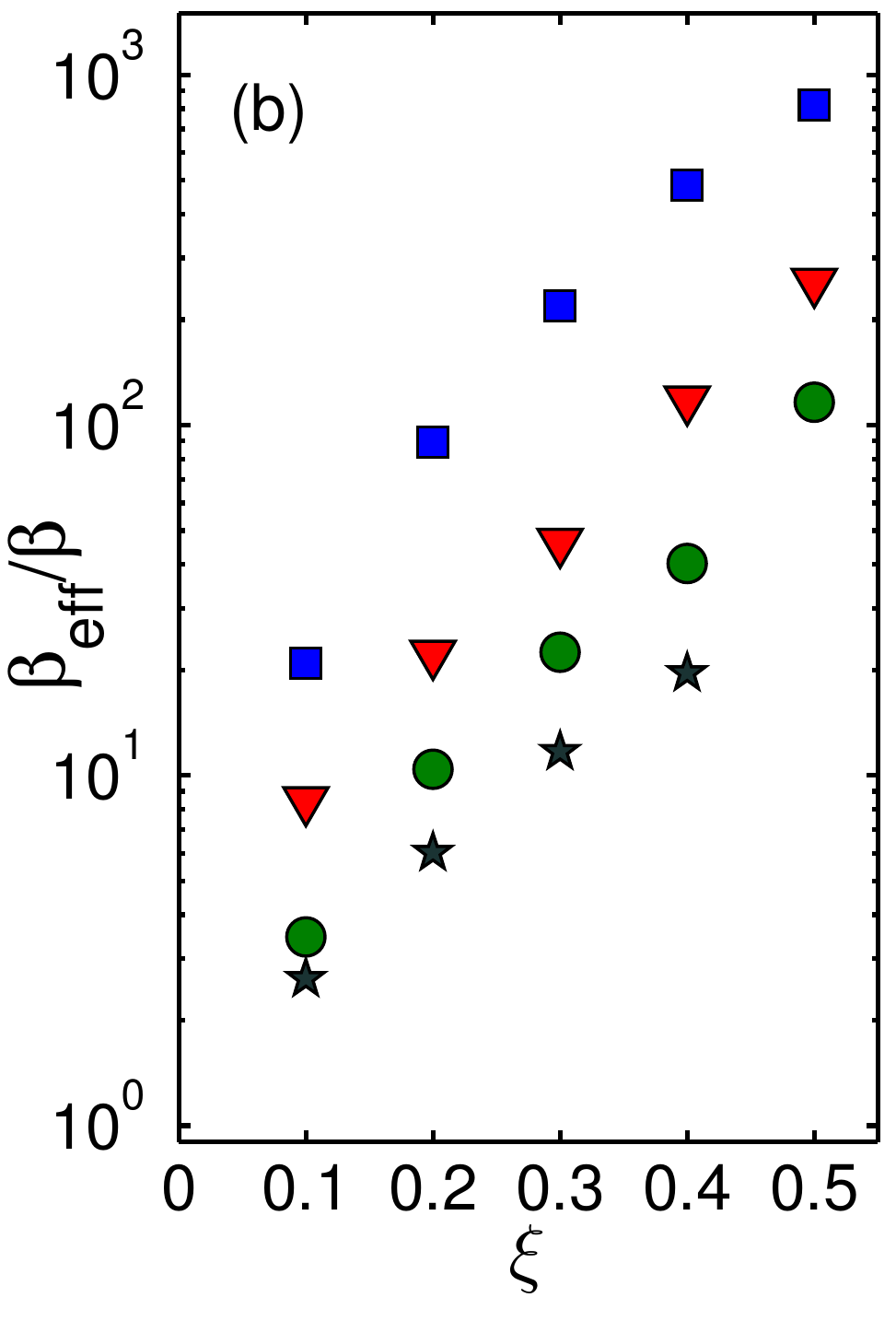}} 
\caption{ (a) 
$Nu$ vs. $Ra$  and (b) $\beta_{eff}/\beta$ vs. $\xi$ for 50000 bubbles. 
In (a), the numerical results are shown as crosses ($\xi=10^{-3}$), squares ($\xi=0.1$), triangles ($\xi=0.2$), circles ($\xi=0.3$), diamonds ($\xi=0.4$) and stars ($\xi=0.5$). 
Simulations without bubbles are also shown for comparison as a dashed line joining small dots and data from the LB simulations of Ref.~\cite{Toschi12} as filled circles, red for no-boiling and blue for boiling
In the inset, the effective scaling exponent $\gamma(\xi)$ obtained from power-law fits of the form $Nu\propto Ra^{\gamma}$ is shown as a function of $\xi$ \textcolor{black}{for 50000 (blue squares) and 150000 (red circles) bubbles}. In (b), the effective buoyancy has been computed from Eq. (1). \textcolor{black}{The symbols are the same as in Figure~\ref{fig:NuvsXi_fixedNb50}a}.}
\end{center}
\label{fig:NuwithRa_fixedNb50}
\end{figure}

The heat transport in  single-phase RB convection can be approximated 
by an effective scaling law $Nu=A_0 \, Ra^{\gamma_0}$. In the  present $Ra$ range the experimental
data are well represented with the choices  $\gamma_0=0.31$ and $A_0\simeq$0.120. How does the 
effective scaling law change for  boiling convection? 
Figure~\ref{fig:NuwithRa_fixedNb50}a shows the Nusselt number vs. Rayleigh number for different values of $\xi$ for $N_b = 50000$ bubbles. The two solid lines have slopes $1/3$ and $1/5$, while the dashed line  shows the single-phase values.  If we fit $Nu$ for the boiling case again with an effective scaling law 
$Nu=A(\xi)Ra^{\gamma(\xi)}$, we obtain the effective exponents $\gamma (\xi) $ shown in the inset of the figure (as blue squares). Of course,  $\gamma (\xi =0) =\gamma_0$ and, as $\xi$ increases, 
$\gamma(\xi) $ decreases to a value close to 0.20. 
In the range $0\leq\xi\leq0.5$ the numerical results for $A(\xi)$ are well represented by $A(\xi)/A_0=1+66.31\,\xi$, which monotonically increases from 1 to 33.15 for $\xi=0.5$. 
How strongly do the pre-factor $A(\xi)$ and the effective scaling exponent $\gamma(\xi)$ depend on $N_b$? 
In the inset of the same figure, we show $\gamma (\xi ) $ for $N_b= 150000$ bubbles (see red-circles),
in order  to compare   with the $N_b= 50000$ case. 
 The functional dependence $\gamma(\xi)$ is very close for the two cases. Further, we find $A(\xi)/A_0=1+83.54\,\xi$ for 150000 bubbles, i.e., a stronger $\xi$ dependence as compared to the $N_b = 50000$ case, reflecting
 the enhanced number of bubbles.  
 
It is tempting to regard the increased heat transport as due to the additional buoyancy provided by the bubbles. 
In this view, the Rayleigh number should be based on an effective buoyancy $\beta_{eff}$ in place of the pure liquid buoyancy $\beta$.  An expression for $\beta_{eff}$ can then be found by equating 
$A(\xi) \, Ra^{\gamma(\xi)}$ to $A_0\, \left[(\beta_{eff}/\beta) \, Ra\right]^{\gamma_0}$ with the result 
\begin{equation}
{\beta_{eff}\over\beta} = \left[A(\xi)\over A_0\right]^{1\over\gamma_0} \, Ra^{{\gamma(\xi)\over\gamma_0}-1}.
\end{equation} 
The quantity $\beta_{eff}/\beta$ as given by this relation is shown in Figure
\ref{fig:NuwithRa_fixedNb50}b 
as function of $\xi$ and $Ra$ for $N_b=50000$. For the same $\xi$, $\beta_{eff}$ decreases as $Ra$ increases as expected on the basis of Figures~\ref{fig:Nuphaseplot} and~\ref{fig:NuvsXi_fixedNb50}. For fixed $Ra$,  $\beta_{eff}/\beta$ increases with $\xi$, also as expected. 
It is quite striking that $\beta_{eff}$ can exceed $\beta$ by nearly 3 orders of magnitude for $\xi=0.5$ and small Rayleigh number. 
Note that one cannot directly compare the numerical values for $\beta_{eff}/\beta$ shown in figure 
\ref{fig:NuwithRa_fixedNb50}b with an experiment in which $\xi$ is increased in a given cell, as in our plot $N_b=50000$ is fixed, whereas in the experiment
$N_b\sim \xi^m$ with $m\simeq 3.4$  as discussed above. 

A recent Lattice-Boltzmann (LB) simulation of finite-size bubbles also found heat transport enhancement~\cite{Toschi12}. The results of this study for $Ra\sim10^7$ are shown by filled circles in Figure~\ref{fig:NuwithRa_fixedNb50}a.
The heat transfer enhancements achieved are much smaller than ours, most likely due to the significantly smaller number of bubbles (only a few hundreds), as well as other differences (the values of $Ja$, $Pr$ etc.) of lesser importance.

\begin{figure}[h!]
\begin{center}
\subfigure{\includegraphics[width=0.23\textwidth]{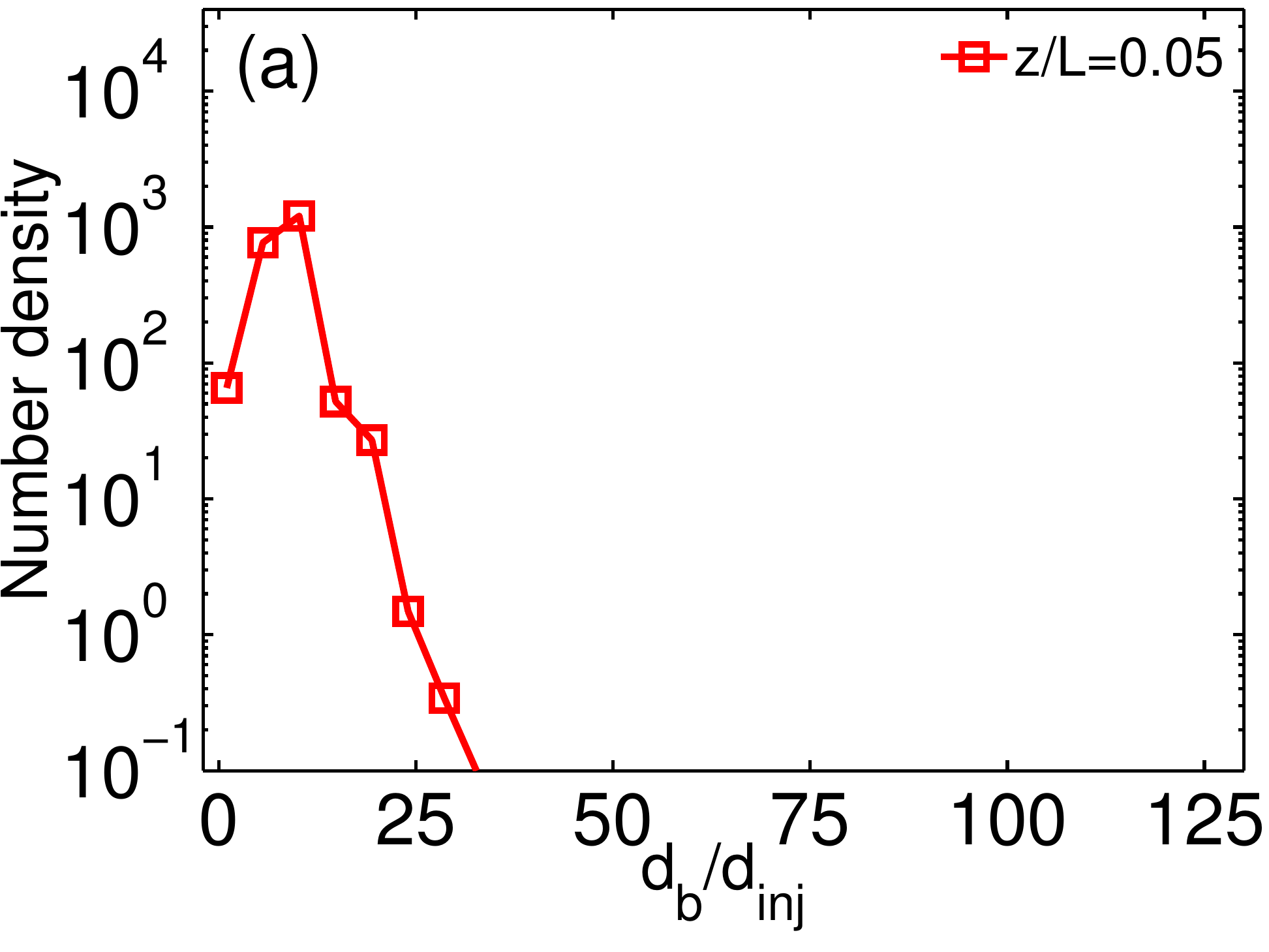}}
\subfigure{\includegraphics[width=0.23\textwidth]{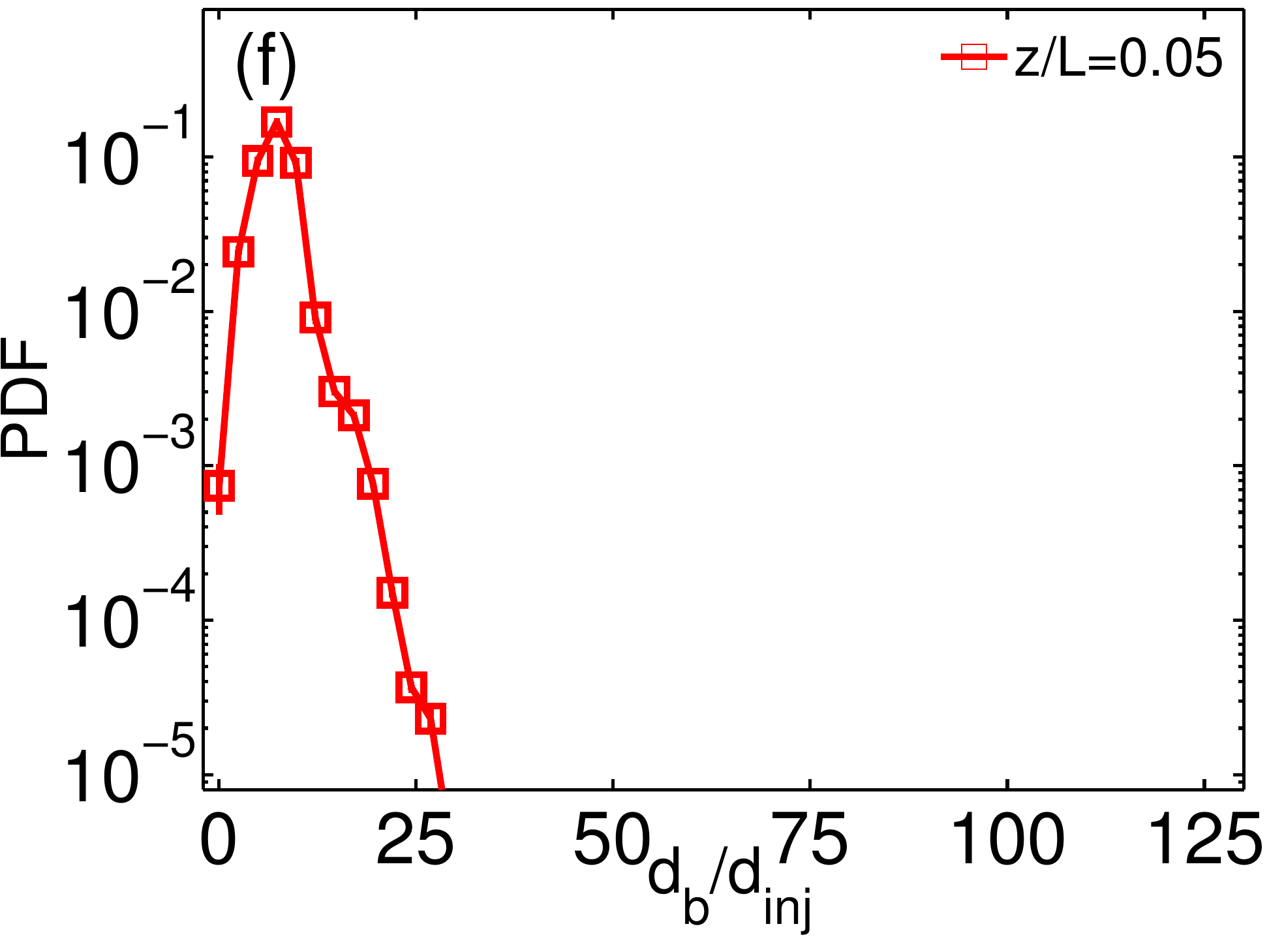}} \\
\vspace{0.1cm}
\subfigure{\includegraphics[width=0.23\textwidth]{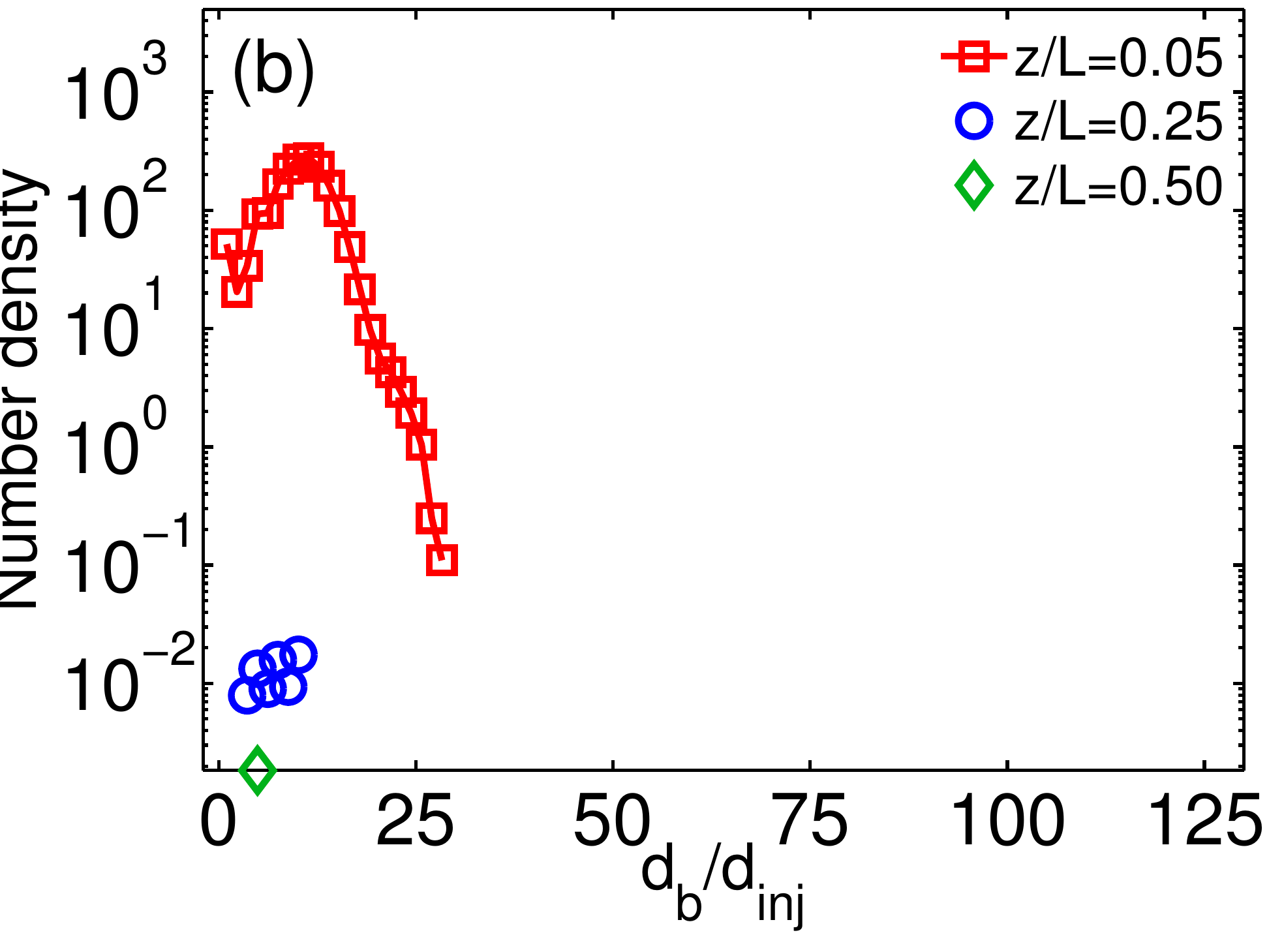}}
\subfigure{\includegraphics[width=0.23\textwidth]{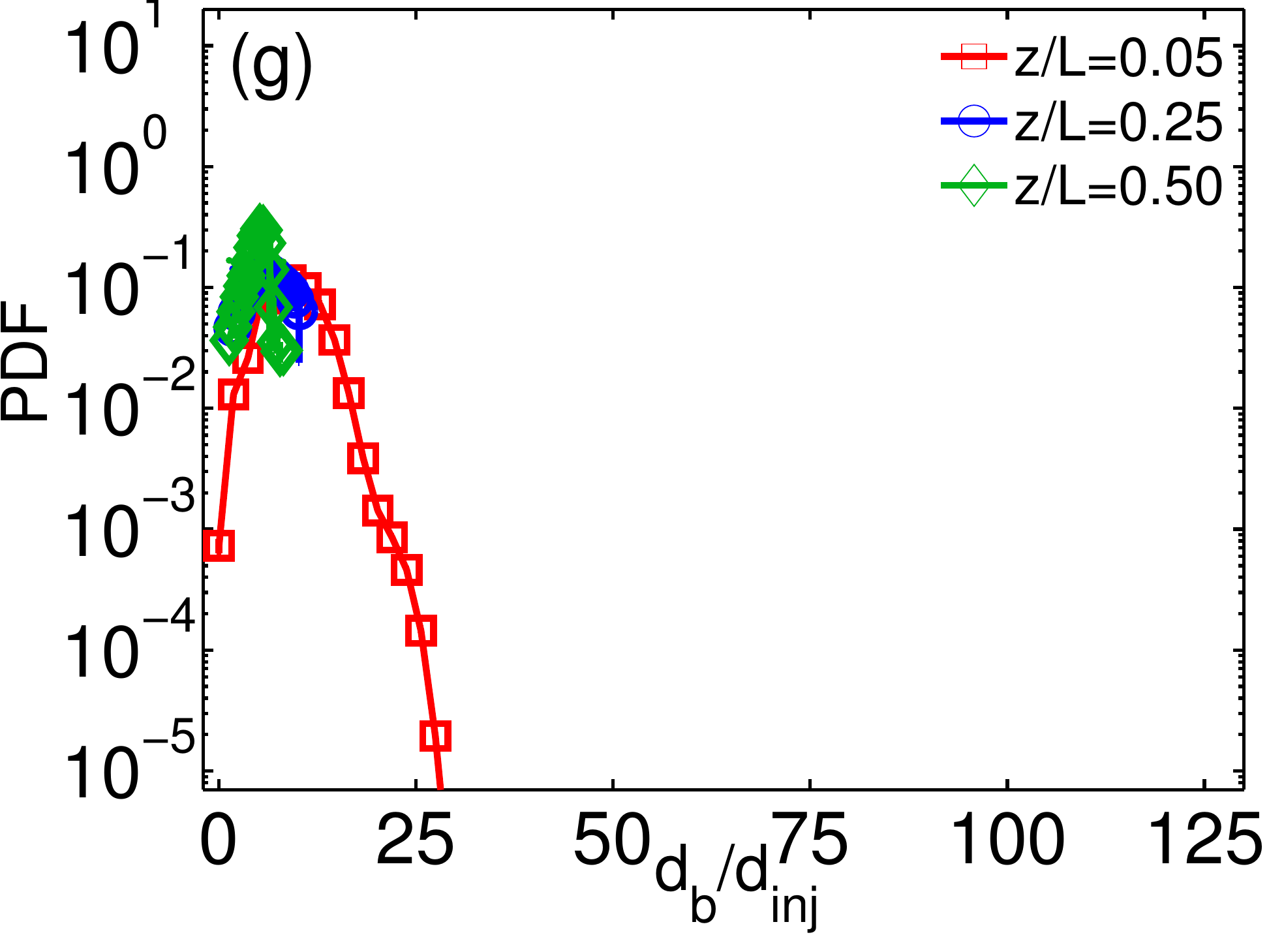}} \\
\vspace{0.1cm}
\subfigure{\includegraphics[width=0.23\textwidth]{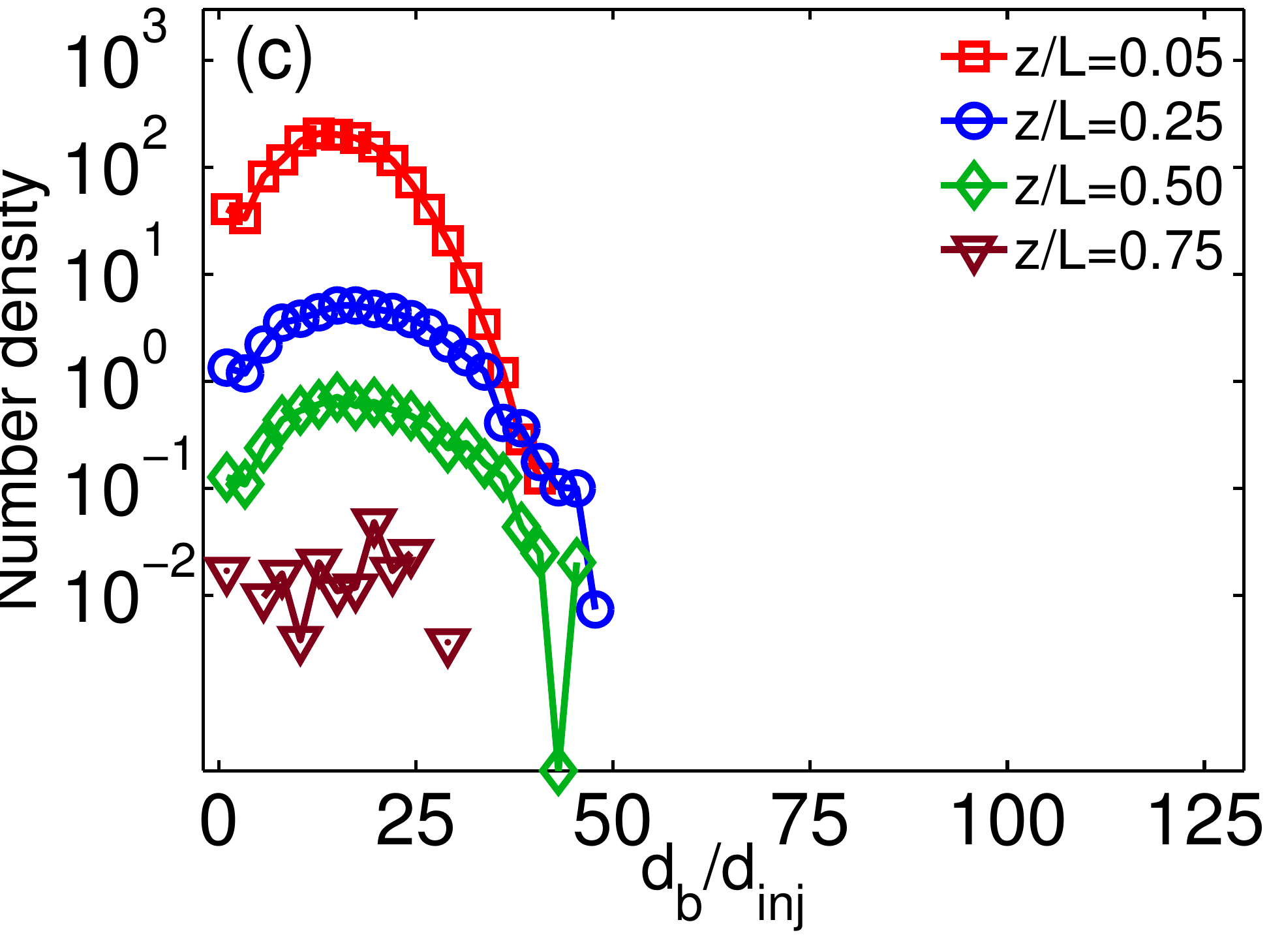}}
\subfigure{\includegraphics[width=0.23\textwidth]{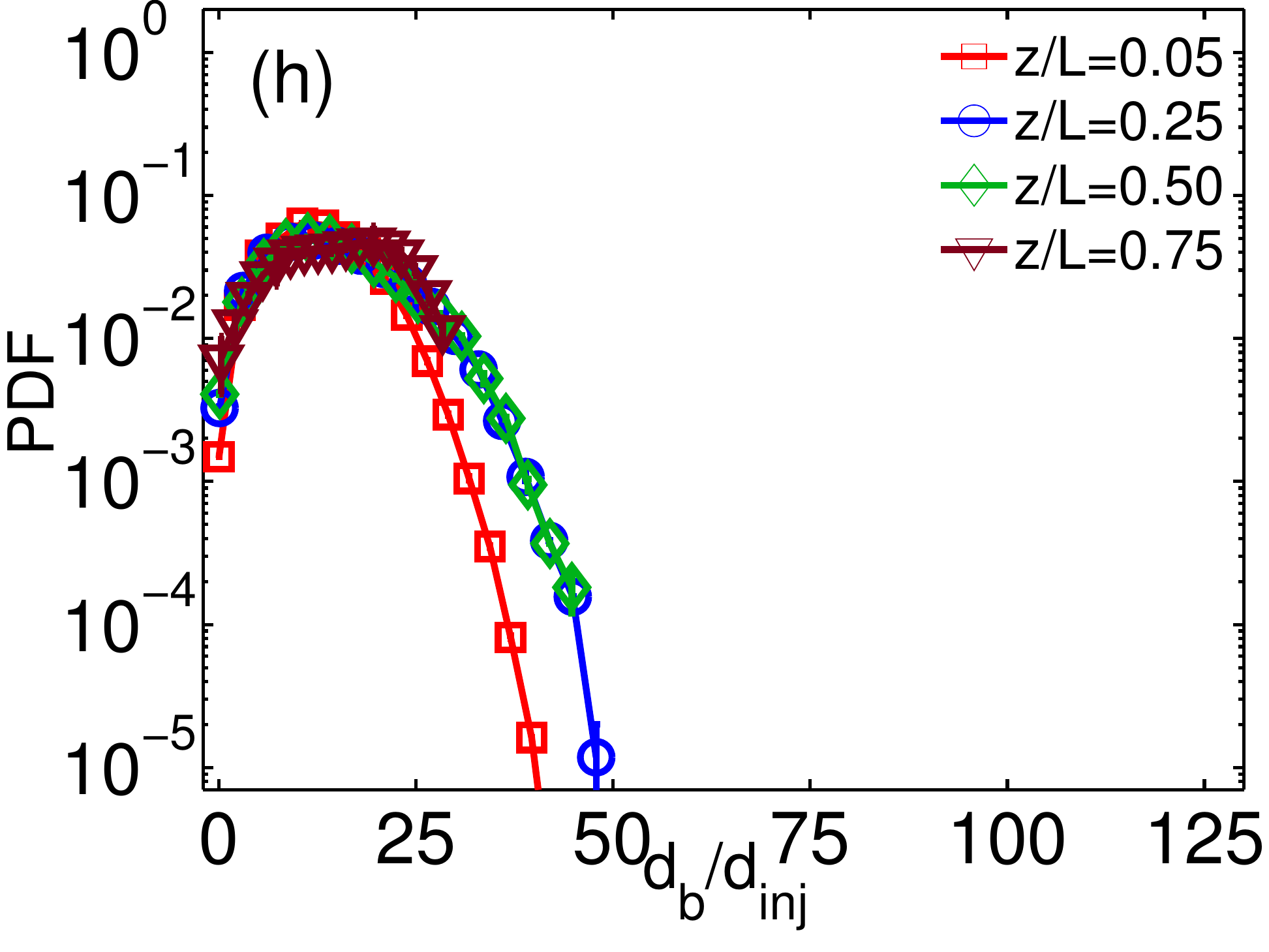}} \\
\vspace{0.1cm}
\subfigure{\includegraphics[width=0.23\textwidth]{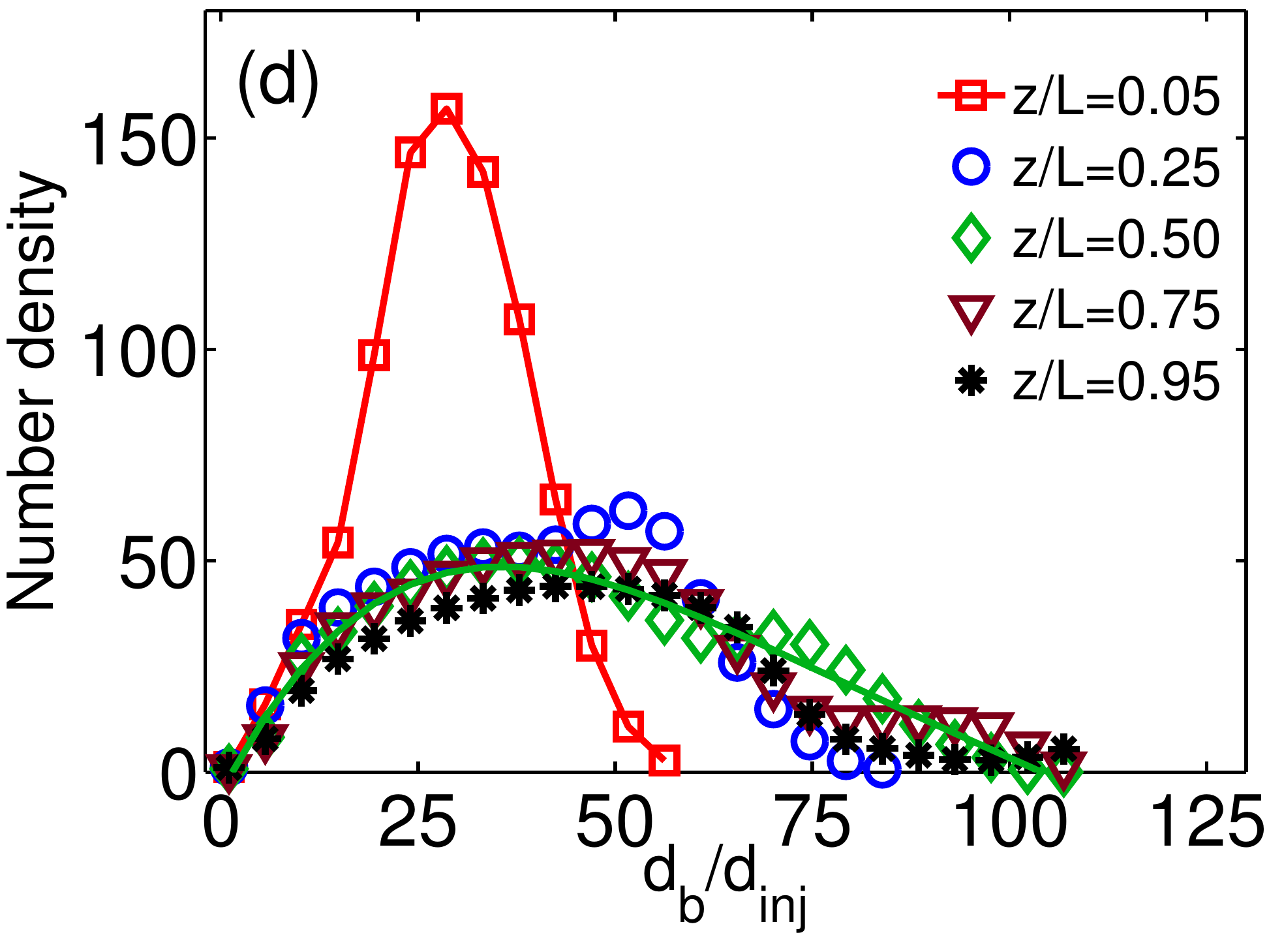}}
\subfigure{\includegraphics[width=0.23\textwidth]{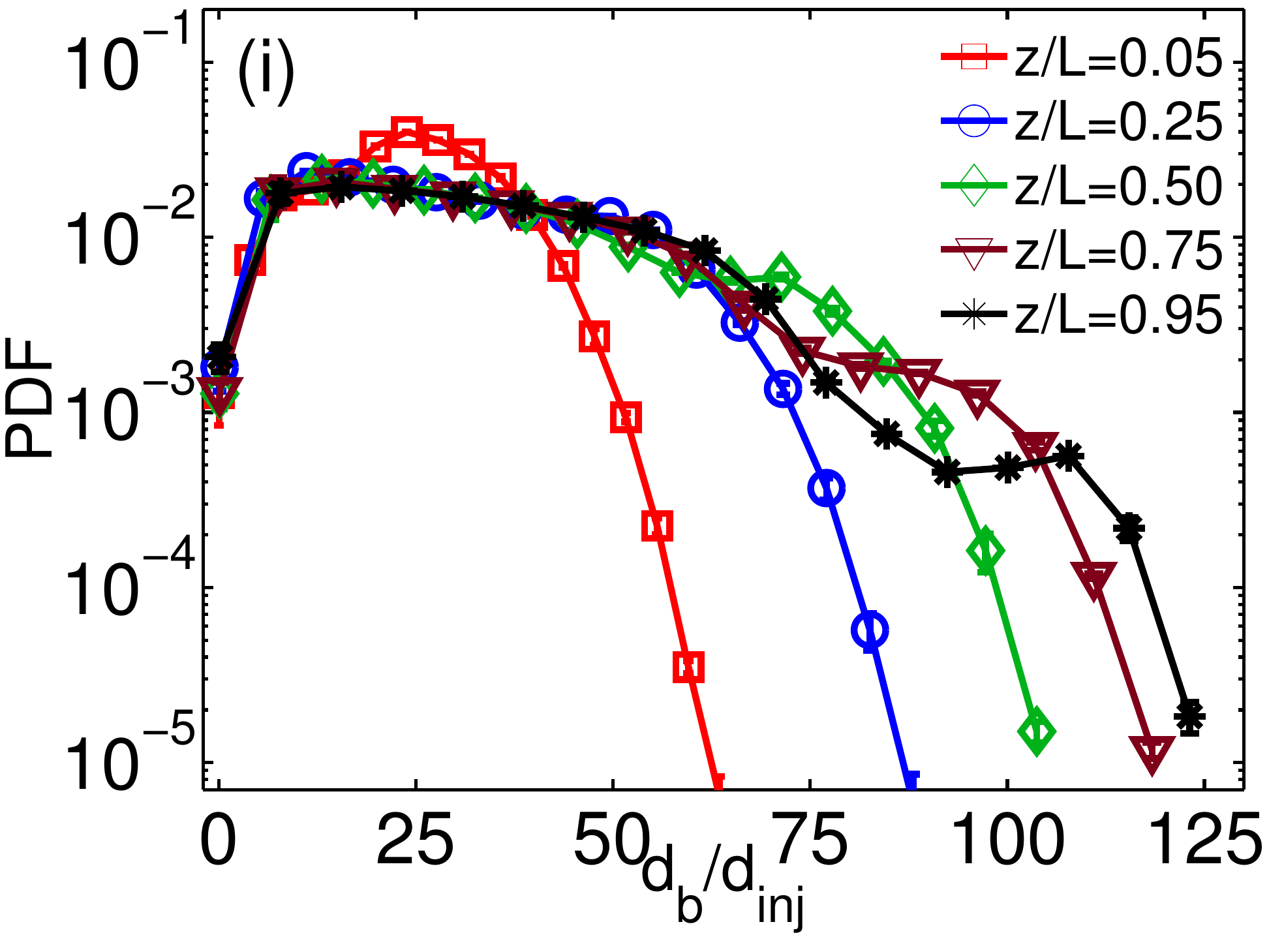}} \\
\vspace{0.1cm}
\subfigure{\includegraphics[width=0.23\textwidth]{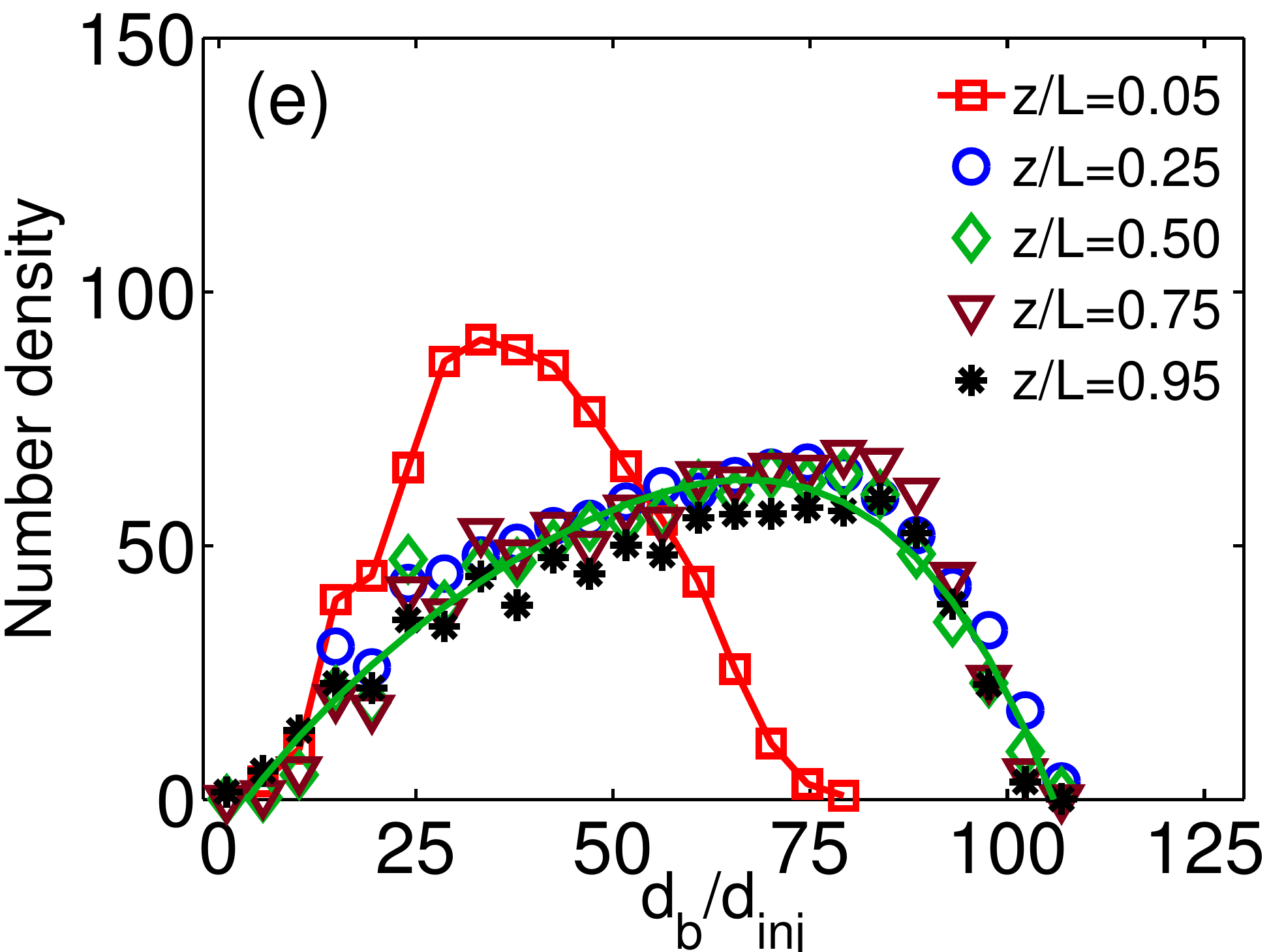}}
\subfigure{\includegraphics[width=0.23\textwidth]{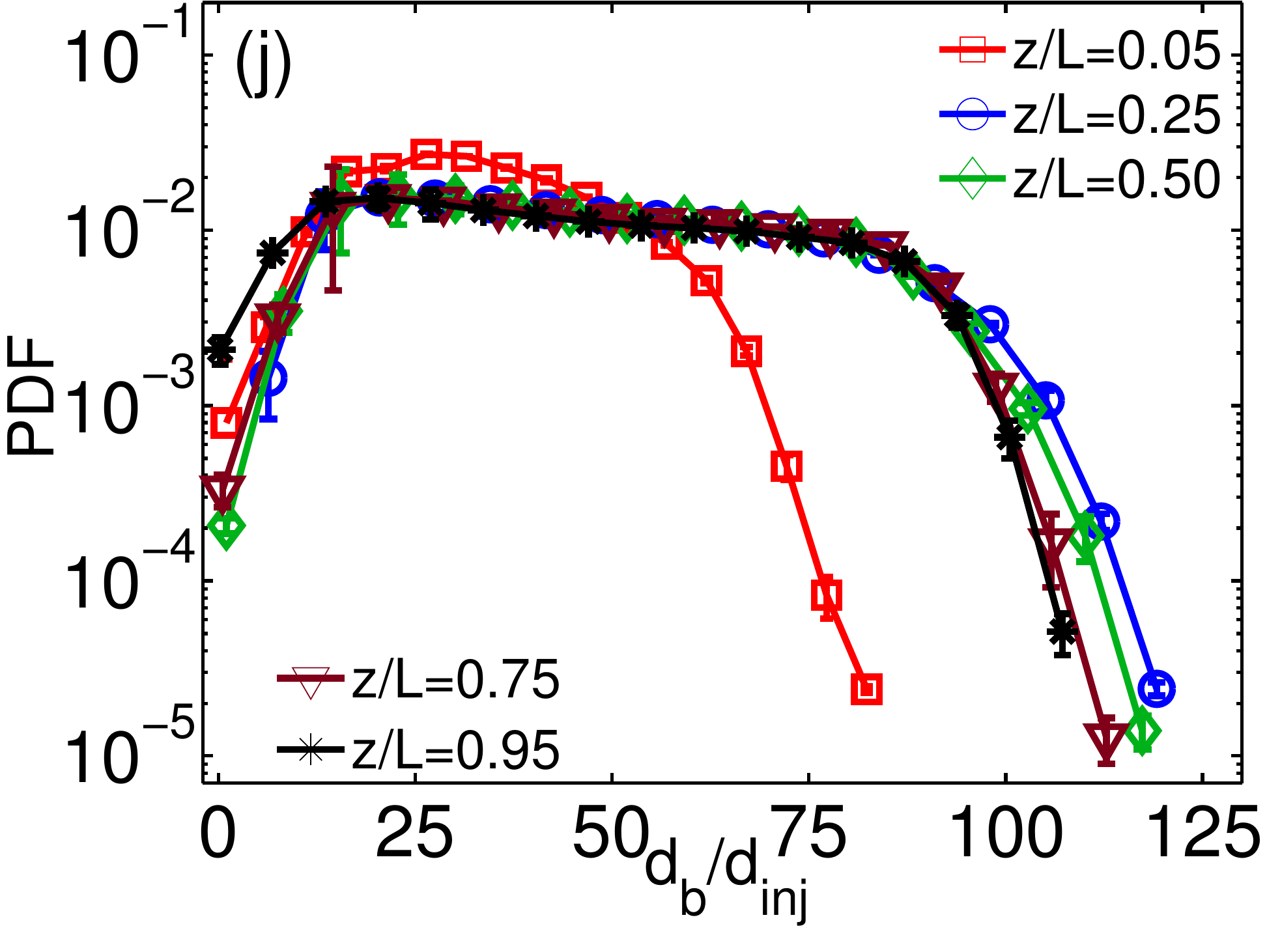}} \\
\vspace{0.1cm}
\caption{Bubble size (given in bubble diameters $d_b/d_{inj}$) distributions 
at various heights in the cylinder: (a-e) bubble density vs. bubble diameter, 
and (f-j) probability density function vs. bubble diameter for Ra$=2\times10^8$ and $N_b=50000$.
From top to bottom, $\xi=$ 0.1 (a+f), 0.2 (b+g), 0.3 (c+h), 0.4 (d+i), and 0.5 (e+j). 
The 
bubble injection diameter at the hot plate is $d_{inj}=$ 38 microns. Data symbols for different vertical heights $z/L=0.05$ (red-squares), 0.25 (blue-circles), 0.5 (green-diamonds), 0.75 (brown-triangles), and 0.95 (black-stars). 
Note that all the vertical axes are in logarithm co-ordinates except figures (d) and (e).
}
\end{center}
\label{fig:bubsize}
\end{figure}

\begin{figure}[h!]
\begin{center}
\subfigure{\includegraphics[width=0.48\textwidth]{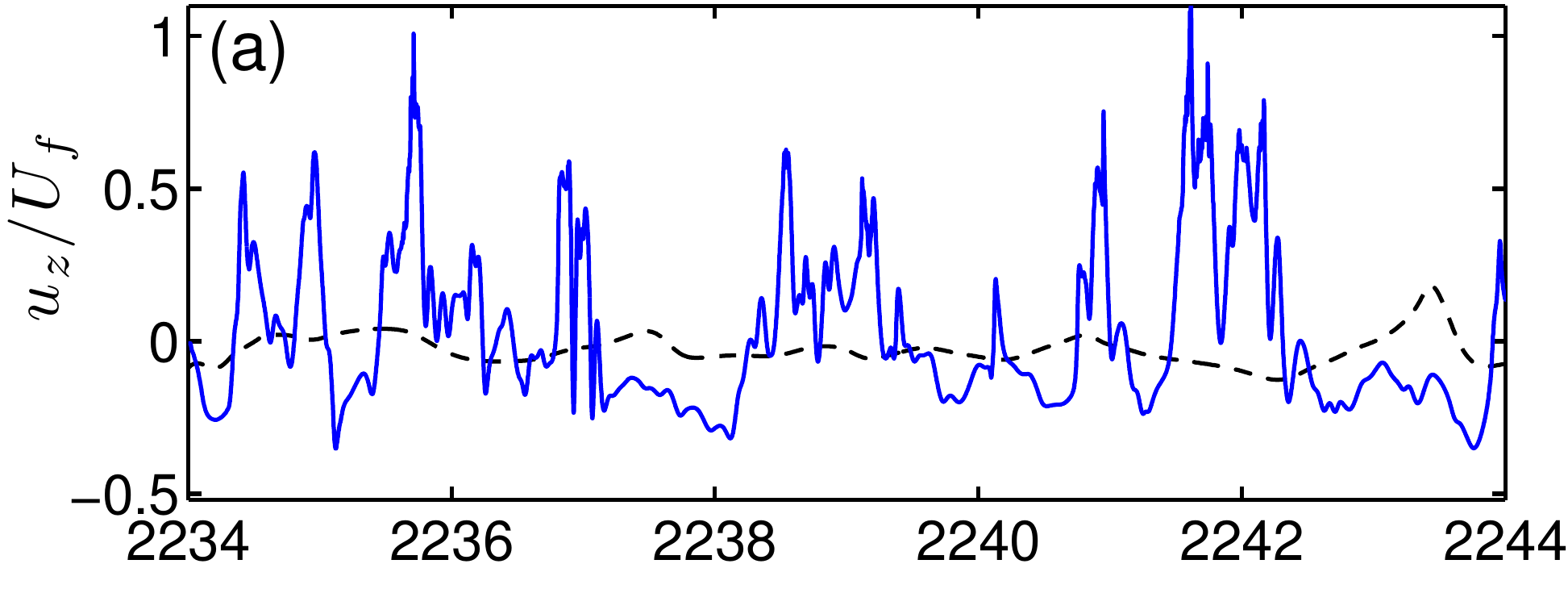}} \\
\subfigure{\includegraphics[width=0.48\textwidth]{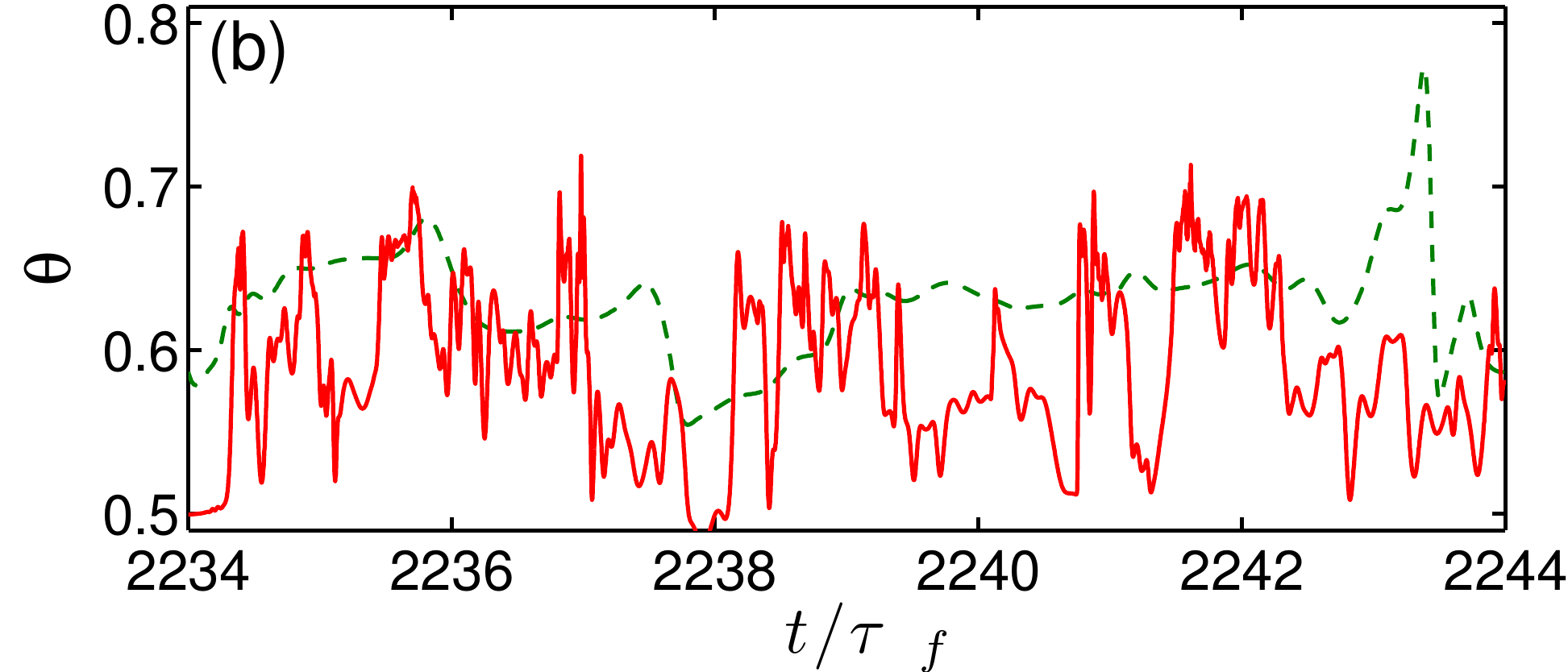}} 
\caption{The solid lines show the dimensionless vertical velocity $u_z/U_f$ (top panel) and temperature $\theta$ (bottom panel) as functions of dimensionless time in the hot liquid at a height $z/L=0.02$ near the axis. The dashed lines show similar results for simulations without bubbles; here $Ra=2\times 10^8$, $N_b=150000$ and $\xi=0.3$.}
\end{center}
\label{fig:probevelocity_Ra2e8}
\end{figure}

\begin{figure}[h!]
\begin{center}
\subfigure{\includegraphics[width=0.23\textwidth]{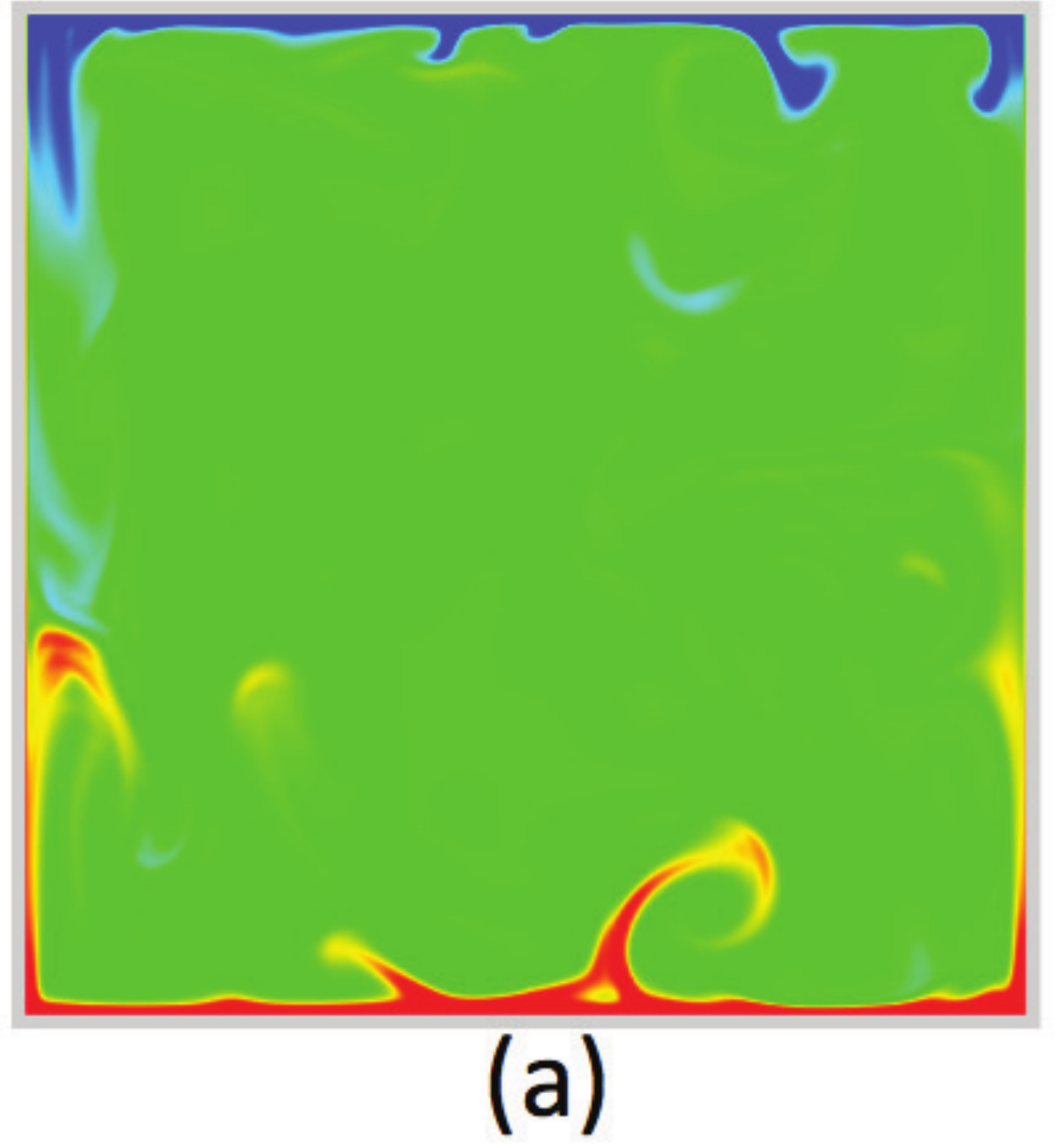}} 
\subfigure{\includegraphics[width=0.23\textwidth]{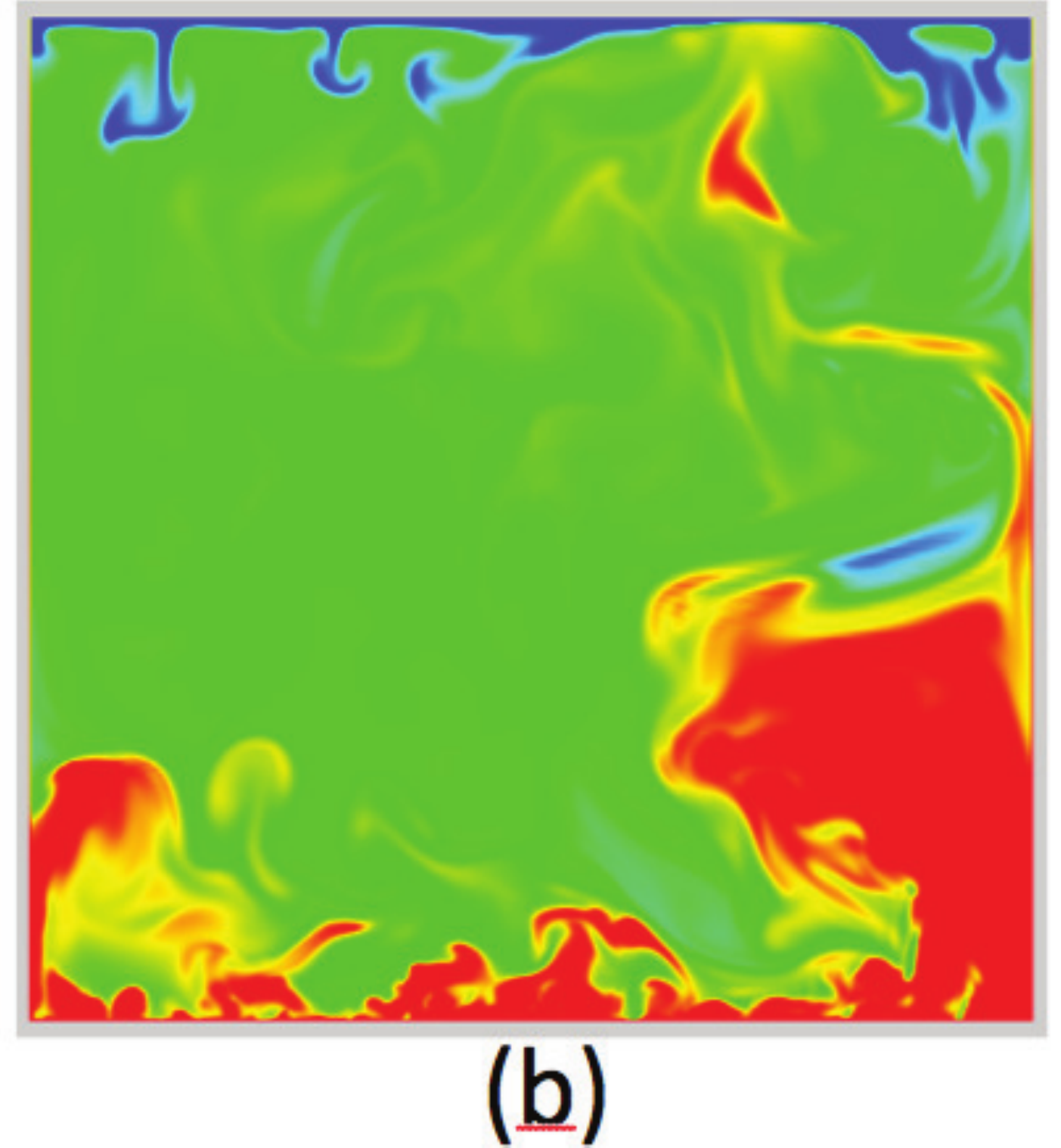}} 
\caption{Instantaneous dimensionless temperature field in a vertical plane through the cell axis for convection without bubbles (left) and with bubbles (right). The color varies from red for $\theta=0.7$ to blue for $\theta=0.3$; here $Ra=2\times 10^8$, $N_b=150000$ and $\xi=0.3$. }
\end{center}
\label{fig:Tsnapshots}
\end{figure}
 
\textcolor{black}{Where are the bubbles in the flow and 
how are they distributed in size, depending on their location? 
In Figure~\ref{fig:bubsize}, the statistics on the bubble diameter computed at different vertical heights in the cylinder are shown. 
To get an immediate impression on how large the  bubbles have  grown, we have normalized the bubble diameter ($d_b$) with the initial injection diameter $d_{inj}=38 \mu m$. We calculate the time averaged  bubble density in thin horizontal 
slices positioned at 5 different vertical heights in the cylinder  for various diameter ranges, see Figure~\ref{fig:bubsize} (a-e). 
For small superheat $\xi=0.1$ the bubble nuclei do not grow much:
 Most of them only up to a diameter 12 times the injection size and only very few towards 25 times the injection
 diameter (see a). Moreover, they do not make it up to one quarter of the cell height, as they 
 encounter cold liquid  and condense. As we increase $\xi$, the bubbles grow to larger sizes 
 and can even reach the top plate (see Figs. ~\ref{fig:bubsize}b to \ref{fig:bubsize}e). 
 Though the number density at a given cross-section decreases
  a wide range of bubble size emerges, leading to poly-dispersity. 
  The bubbles can grow up to a size of even 100 times the initial injection diameter. 
  Note that for large $\xi=0.4$ and even more at $\xi=0.5$, at any plane away from the boundary layers the number density shows a similar trend for bubble size distribution, reflecting the homogeneously boiling situation. In the right column of Figure~\ref{fig:bubsize} (see f-j), we show the corresponding 
   probability density functions (PDFs) vs. the bubble diameter, now all in log-linear form. 
   Again we see that both the bubble maximum and the most probable diameter increase
    as we increase $\xi$.}

\begin{figure}[h!]
 \begin{center}
\subfigure{\includegraphics[width=0.5\textwidth]{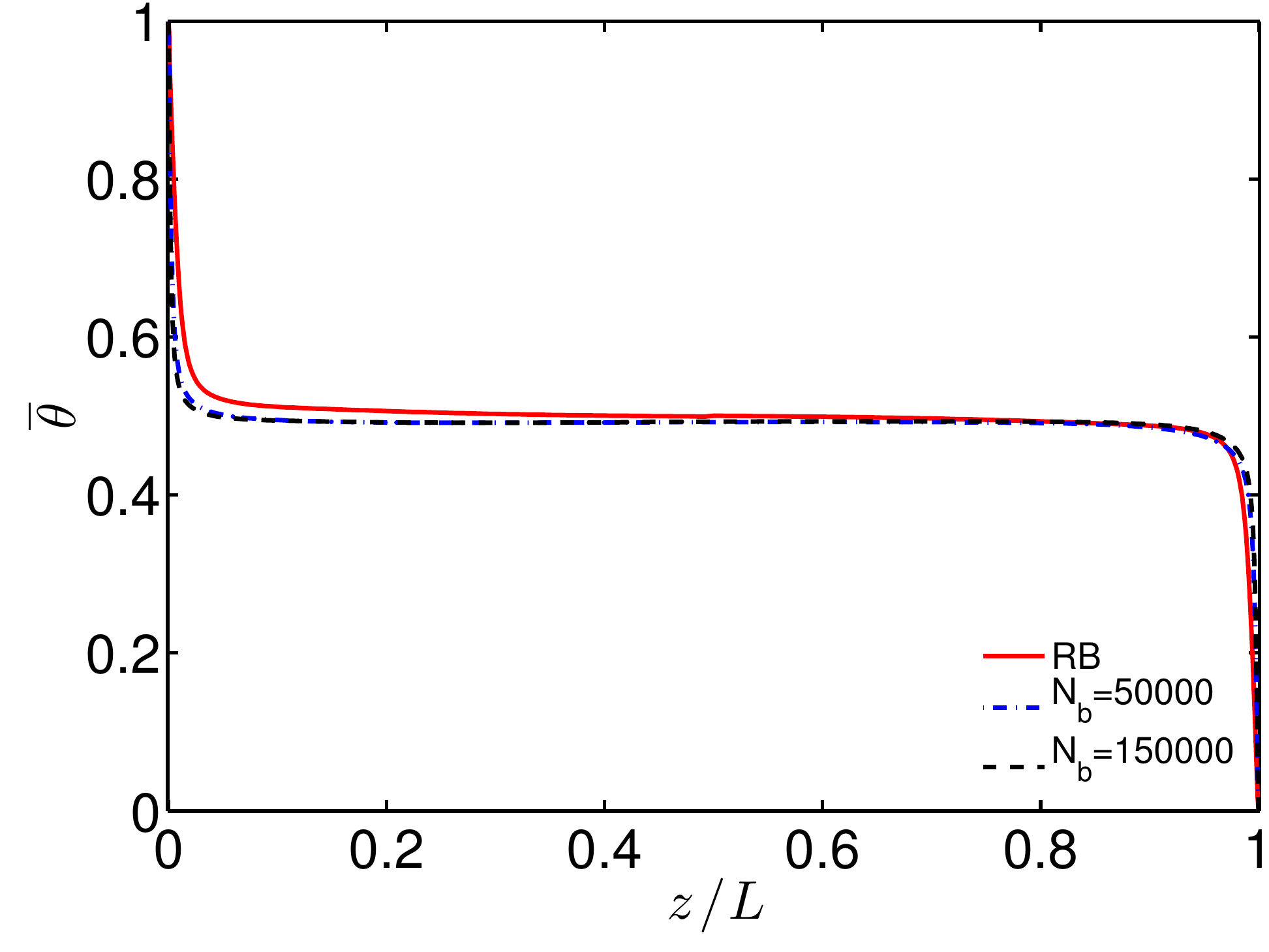}} 
\caption{Normalized temperature $\overline{\theta}$ vs. $z/L$ over the entire cell height. 
Here $\xi=0.5$, $Ra=2\times10^8$ and RB (continuous red), $N_b=50000$ (dash-dot blue), and $150000$ (dash black). }
\end{center}
\label{fig:CompleteTemp}
\end{figure}

We now come to the local flow organization. As well known 
the boundary layers formed on the bottom (and top) plate are marginally stable and occasional intermittent eruptions of hot (or cold) liquid occur at their edges. 
Vapor bubbles subject these boundary layers to intense fluctuations which enhance the convective effects.
As an example, Figure~\ref{fig:probevelocity_Ra2e8} shows sample time records of the dimensionless vertical velocity $u_z/U_f$ (top panel),  and temperature $\theta=(T-T_c)/\Delta$ (bottom panel) vs. normalized time $t/\tau_f$ near the axis at $z/L=0.02$, i.e., just outside the hot thermal boundary layer. The velocity scale $U_f$ is defined by $U_f=\sqrt{g\beta\Delta L}$ and  $\tau_f=L/U_f$.
The dashed lines are results for the single-phase case. 
The immediate observation is that the small-scale fluctuations are much stronger in the two-phase case. 
As expected, the positive and negative velocity fluctuations are correlated with warm and cold temperature fluctuations, respectively.

To give an impression of the difference brought about by the presence of bubbles on the convective motions in the cell, we show in Figure~\ref{fig:Tsnapshots} snapshots of the dimensionless temperature in a vertical plane through the axis of the cell for $Ra=2\times 10^8$ in the single-phase (a) and two-phase (b) cases, the latter for $\xi=0.3$ and $N_b=150000$. We notice that bubbles considerably thicken the layer of hot fluid near the base and make it more energetic compared to the single-phase situation. Chunks of hot liquid can be seen all the way up near the cold plate, presumably caused by the latent heat deposited by condensing bubbles in the bulk liquid. The up-down symmetry of the single-phase case that can be seen in the left panel is markedly absent in the two-phase case because of the tendentially upward motion of the bubbles which condense on encountering liquid colder than $T_{sat}$. 
This mechanism is evidently quite different from the symmetry-breaking process observed in non-Boussinesq systems which is due to the temperature dependence of viscosity, see e.g., Ref.~\cite{sugiyama08}. 

\begin{figure}[h!]
 \begin{center}
\subfigure{\includegraphics[width=0.48\textwidth]{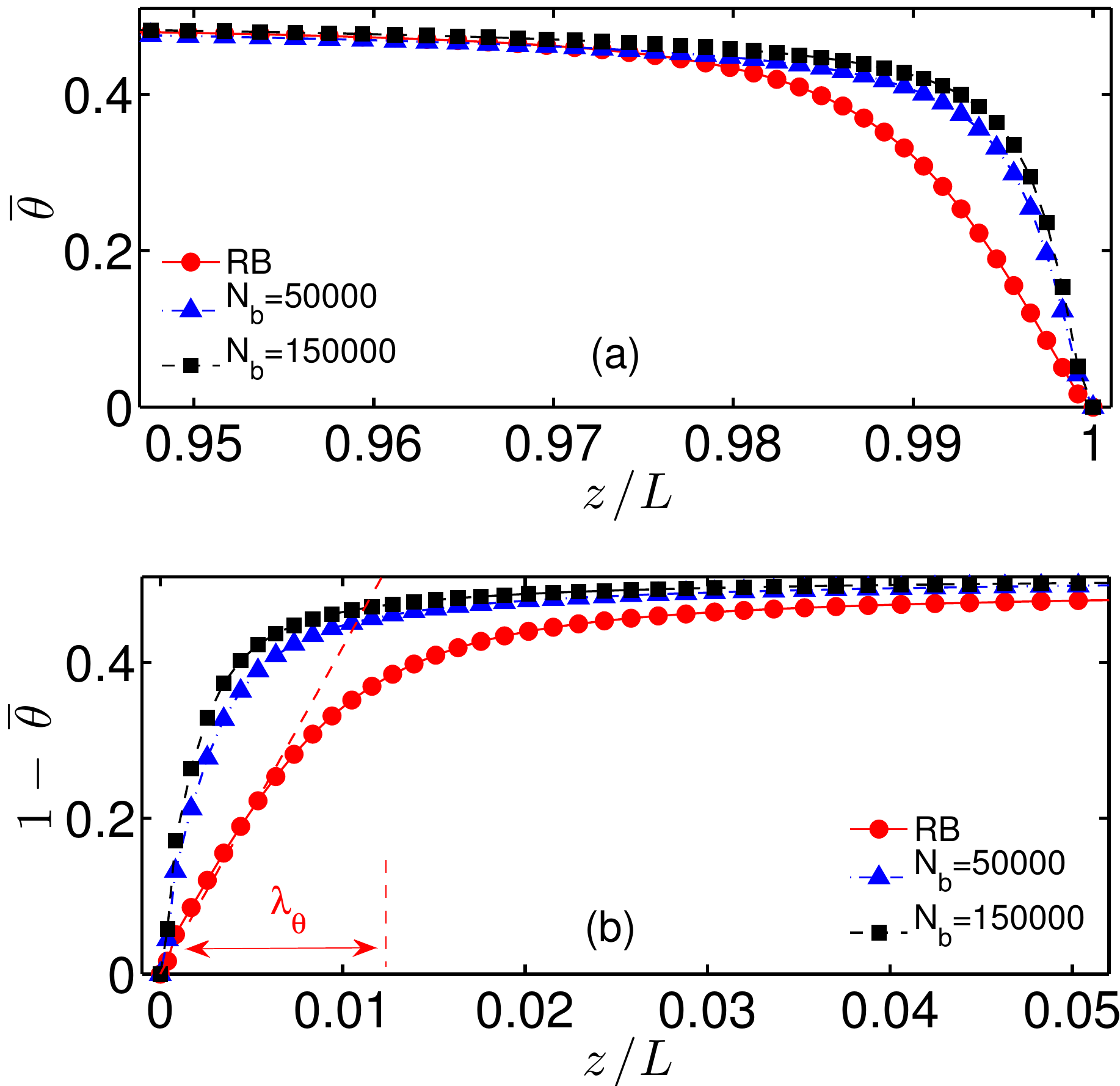}} 
\caption{(a) $\overline{\theta}$ vs. $z/L$ near cold plate, and (b) $1-\overline{\theta}$ vs. $z/L$ near hot plate.
Symbols are red circles (RB), blue triangles ($N_b=$50000) and black squares ($N_b=$150000). 
Boundary layer thickness $\lambda_{\theta}$ based on the wall gradient is also indicated in (b) for single-phase convection. 
Here $\xi=0.5$ and $Ra=2\times 10^8$.}
\end{center}
\label{fig:boundary_coldhot}
\end{figure}

Figure~\ref{fig:CompleteTemp} shows the time and area-averaged mean temperature in the cell as a function of height. The red line is the single-phase case, while the dashed line is for the two-phase cases with $\xi=1/2$; the results for the different bubble numbers superpose within the resolution of this figure. For both cases the dimensionless bulk mean temperature is close to 0.5, i.e.,  $T\simeq T_m={(T_h+T_c)/2}=T_{sat}$. A close inspection of the figure shows that this tendency is enhanced in the two-phase case because the bubble surface temperature is fixed at the same value $T_{sat}$.  It may be expected that, in this situation,  not many bubbles will grow or condense in the bulk and only condense once they reach the neighborhood of the top plate. The figure also shows that, in the two-phase case, the boundary layers near the top and bottom plates are not symmetric. This feature can be seen more clearly in Figure~\ref{fig:boundary_coldhot} where details of the mean temperature near the upper and lower plates are shown. The lack of symmetry between the two is more evident here together with the thinning of the boundary layers with increasing numbers of bubbles. This latter result is a clear manifestation of the enhanced convective circulation inside the cell promoted by increasing the number of bubbles.

\begin{figure}[h!]
 \begin{center}
\subfigure{\includegraphics[width=0.43\textwidth]{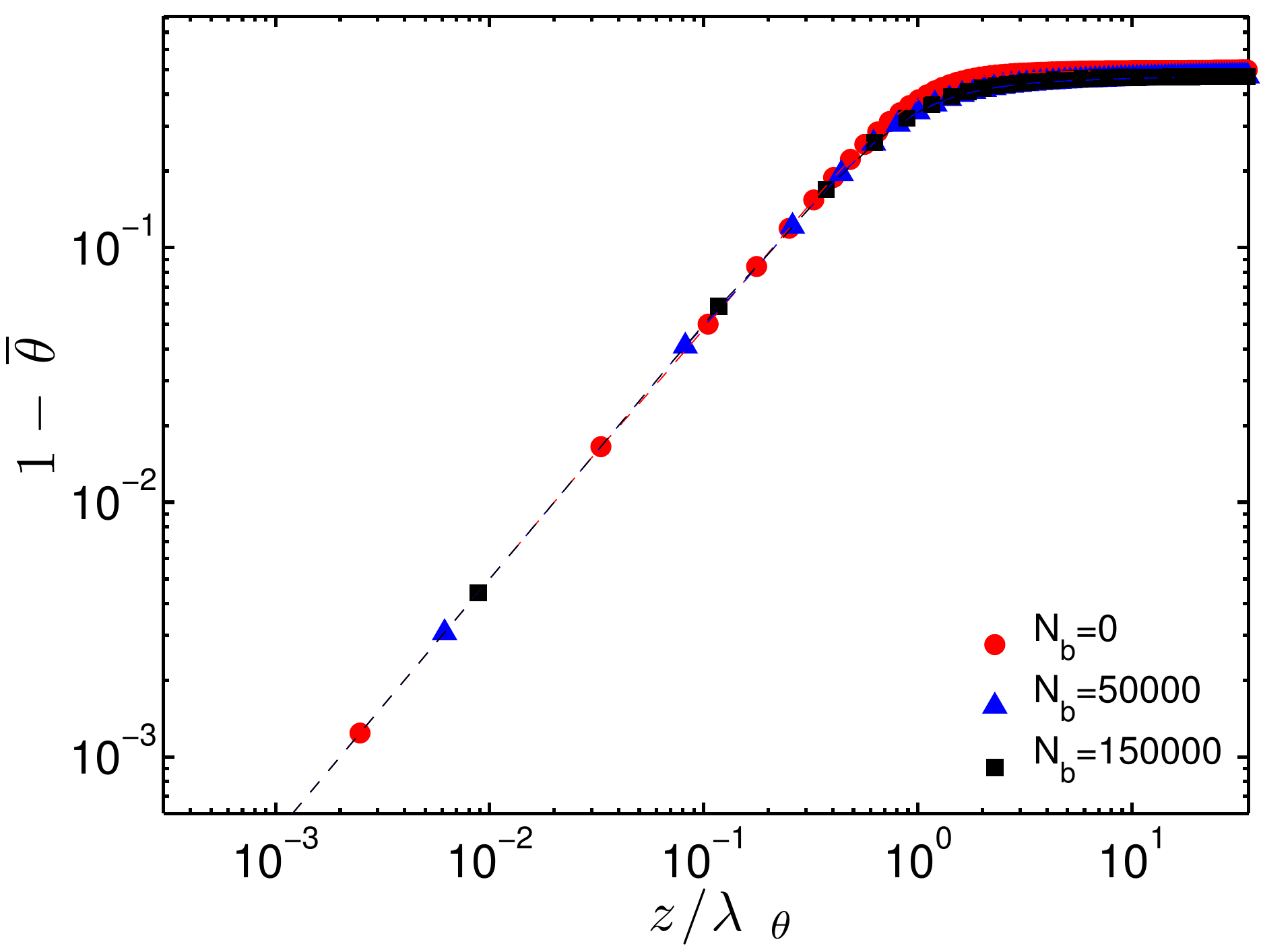}} 
\caption{Normalized mean temperature $1-\overline{\theta}$ vs. $z/\lambda_{\theta}$ in the hot thermal boundary layer. 
Here $\xi=0.5$, $Ra=2\times 10^8$ and $N_b=0$ (red circles), 50000 (blue triangles) and 150000(black squares). 
}
\end{center}
\label{fig:boundarylayer2}
\end{figure}

For the hot plate, one can define the thermal boundary layer thickness as $\lambda_{\theta}=(T_m-T_h)/[dT/dz]_{z=0}$ where $[dT/dz]_{z=0}$ is the mean temperature gradient at the hot plate. Replotting the data of the bottom panel in Figure~\ref{fig:boundary_coldhot} as functions of $z/\lambda_{\theta}$, we find the results shown in Figure~\ref{fig:boundarylayer2}. 
The three sets of data now collapse on a single line in the range $0\le z/\lambda_{\theta}\le 0.5$.
The small differences farther away from the wall reflect differences in the shape factor of the boundary layers. 

\section{Summary and Conclusions}

In summary, our investigation of a simple model of Rayleigh-B\'{e}nard convection with boiling has demonstrated the effect of the degree of superheat and of the bubble number on heat transport. Comparison with existing data suggests a basic conformity of our results with some physical features of a real systems. Vapor bubbles significantly enhance the heat transport primarily by increasing the strength of the circulatory motion in the cell. 
The velocity and thermal fluctuations of the boundary layers are increased and, by releasing their latent heat 
upon condensation in the bulk fluid, the bubbles also act as direct carriers of energy. 
We have shown that the heat transfer enhancement can be interpreted in terms of an enhanced 
buoyancy which is shown in Eq. (1) and Figure~\ref{fig:NuwithRa_fixedNb50}b. The relative effect of the bubbles diminishes as $Ra$ increases.

\begin{acknowledgments}
We are thankful to G. Ahlers and F. Toschi for providing their valuable data and numerical results and to
L. Biferale and C. Sun for helpful discussions. We acknowledge financial support by the Foundation for Fundamental Research on Matter (FOM) and the National Computing Facilities (NCF) sponsored by NWO. Computations have been performed on the Huygens cluster of SARA in Amsterdam. This research is part of the FOM Industrial Partnership Program on Fundamentals of Heterogeneous Bubbly Flows. P.O. gratefully acknowledges support from FIRB under grant RBFR08QIP5\_001.
\end{acknowledgments}



\end{article}

\begin{thebibliography}{}

\bibitem{dhir98}
V. K. Dhir, 
{\em Boiling heat transfer}, 
Annu. Rev. Fluid Mech. {\bf 30},  365 (1998).

\bibitem{carey1992}
V. P. Carey (1992), {\em Liquid-vapor phase-change phenomena}, 
(Hemisphere, New York, United States).

\bibitem{lienhard2011}
John H. Lienhard IV, and John H. Lienhard V {\em A Heat Transfer Textbook}, 
(Dover Publications, New York, United States).

\bibitem{prosperetti2004}
(Eds.) S. Balachandar, and Andrea Prosperetti, 
{\em IUTAM Symposium on Computational Approaches to Multiphase Flow}, 
(Springer 2004).

\bibitem{prosperetti2007}
(Eds.) Andrea Prosperetti, and Gr{\'e}tar Tryggvasson
{\em Computational Methods For Multiphase Flow}, 
(Cambridge University Press 2007).

\bibitem{lohsermp}
G. Ahlers, S. Grossmann, and D. Lohse, 
{\em Heat transfer and large scale dynamics in turbulent Rayleigh-B\'{e}nard convection}, 
Rev. Mod. Phys. {\bf 81},  503  (2009).

\bibitem{lohsearfm}
D. Lohse and K. Q. Xia, 
{\em Small-scale properties of turbulent Rayleigh-B\'{e}nard convection}, 
Ann. Rev. Fluid Mech. {\bf 42},  335 (2010).

\bibitem{zhong09}
J. Q. Zhong, D. Funfschilling, and G. Ahlers, 
{\em Enhanced heat transport by turbulent two-phase Rayleigh-B\'{e}nard convection}, 
Phys. Rev. Lett. {\bf 102}, 124501  (2009).

\bibitem{oresta09}
P. Oresta, R. Verzicco, D. Lohse, and A. Prosperetti, 
{\em Heat transfer mechanisms in bubbly Rayleigh-B\'enard convection}, 
Phys. Rev. E {\bf 80}, 026304  (2009).
  
\bibitem{Laura11}
L. E. Schmidt, P. Oresta, F. Toschi, R. Verzicco, D. Lohse, and A. Prosperetti,
{\em Modification of turbulence in Rayleigh-B\'enard convection by phase change}, 
  New Journal of Physics {\bf 13},  025002  (2011).  
  
\bibitem{Lakkaraju11}
R. Lakkaraju, L. E. Schmidt, P. Oresta, F. Toschi, R. Verzicco, D. Lohse, and A. Prosperetti, 
{\em Effect of vapor bubbles on velocity fluctuations and dissipation rates in bubbly Rayleigh-B\'enard convection}, 
  Phys. Rev. E {\bf 84},  036312 (2011).    
  
\bibitem{Lakkaraju12}
R. Lakkaraju, R. J. A. M. Stevens, R. Verzicco, S. Grossman, A. Prosperetti, C. Sun, and D. Lohse, 
{\em Spatial dependence of fluctuations and flux in turbulent Rayleigh-B\'{e}nard convection}, Physical Review E, 86, 056315 (2012)
        
\bibitem{richard10}
R. J. A. M. Stevens, R. Verzicco, and D. Lohse, 
{\em Radial boundary layer structure and Nusselt number in Rayleigh-B\'{e}nard convection}, 
J. Fluid Mech. {\bf 643},  495 (2010).

\bibitem{verzicco03}
R. Verzicco and R. Camussi, 
{\em Numerical experiments on strongly turbulent thermal convection in a slender cylindrical cell}, 
J. Fluid Mech. {\bf 477}, 19  (2003).

\bibitem{Toschi12}
L. Biferale, P. Perlekar, M. Sbragaglia, and F. Toschi, 
{\em Convection in multiphase fluid flows using lattice boltzmann methods}, 
Phys. Rev. Lett. {\bf 108},  104502 (2012).

\bibitem{sugiyama08} 
G. Ahlers, E. Brown, F. F. Araujo, D. Funfschilling, S. Grossmann, and D. Lohse, 
{\em Non-Oberbeck-Boussinesq effects in strongly turbulent Rayleigh-B\'{e}nard convection},  
J. Fluid Mech. {\bf 569}, 409-445 (2006). 

\end{thebibliography}
\end{document}